\documentclass[12pt]{article}
\usepackage{hyperref}
\hypersetup{
colorlinks=true,
linkcolor=black,
citecolor=black,
urlcolor=blue,
filecolor=black,
pdfborder={0 0 0},
}

\def\Black{} 
\usepackage{color}
\definecolor{darkblue}{rgb}{0,0,0.5}
\definecolor{darkred}{rgb}{0.5,0,0}
\definecolor{darkgreen}{rgb}{0,0.5,0}
\usepackage{hyperref}
\hypersetup{
colorlinks=true,
linkcolor=black,
citecolor=black,
urlcolor=darkblue,
filecolor=black,
pdfborder={0 0 0},
}
\usepackage{amssymb}
\usepackage{amsfonts}
\usepackage{mathrsfs}
\usepackage{latexsym}
\usepackage{amsmath}
\usepackage{graphics}
\usepackage{graphicx}
\usepackage{sectsty}
\usepackage{setspace}
\usepackage{multirow}
\sectionfont{\normalsize\bf}
\subsectionfont{\rm\normalsize}
\def\be{\begin{equation}}
\def\ee{\end{equation}}
\input amssym.def
\input amssym
\def\Z{{\cal Z}}

\def\P{{\cal P}}

\def\cL{{\cal L}}

\def\T{{\cal T}}

\def\tr{\rm{tr}}

\newcount\nx
\newcount\nxo
\newcount\ny
\newcount\nyo
\newcount\nxd
\newcount\nxod
\newcount\nyd
\newcount\nyod
\newcount\nl

\def\grid#1#2#3#4{
\nxo=#1
\nyo=#2
\nx=#3
\ny=#4    
\multiply \nxo by 4
\multiply \nyo by 4
\advance \nx by 1
\multiply \ny by 4
\linethickness{0.075mm}
\multiput(\nxo,\nyo)(4,0){\nx}{\line(0,1){\ny}}
\nx=#3
\ny=#4
\advance \ny by 1
\multiply \nx by  4
\multiput(\nxo,\nyo)(0,4){\ny}{\line(1,0){\nx}}
}

\def\dgrid#1#2#3#4{
\nxo=#1
\nyo=#2
\nx=#3
\ny=#4
\multiply \nxo by 4
\multiply \nyo by 4
\advance \nxo by - 2
\advance \nyo by - 2
\advance \nx by 1
\multiply \ny by  4
\linethickness{0.3mm}
\multiput(\nxo,\nyo)(4,0){\nx}{\line(0,1){\ny}}
\nx=#3
\ny=#4
\advance \ny by 1
\multiply \nx by  4
\multiput(\nxo,\nyo)(0,4){\ny}{\line(1,0){\nx}}
}

\def\puti#1#2#3{
\nxo=#1
\nyo=#2
\nl=4
\multiply \nxo by \nl
\multiply \nyo by \nl
\divide \nl by 2
\advance \nxo by - \nl
\advance \nxo by - 1
\advance \nyo by - \nl
\advance \nyo by - 1
\put(\nxo,\nyo){$#3$}
}

\def\putl#1#2#3{
\nxo=#1
\nyo=#2
\nl=4
\multiply \nxo by \nl
\multiply \nyo by \nl
\divide \nl by 2
\advance \nxo by - \nl
\advance \nyo by - \nl
\put(\nxo,\nyo){$#3$}
}
\def\puto#1#2{
\nxo=#1
\nyo=#2
\nl=4
\thicklines
\multiply \nxo by \nl
\multiply \nyo by \nl
\divide \nl by 2
\advance \nxo by - \nl
\advance \nyo by - \nl
\put(\nxo,\nyo){\circle{1.5}}
}
\def\case#1#2#3#4{
\nxo=#1
\nyo=#2
\nx=#3
\ny=#4
\nl=4
\multiply \nxo by \nl
\multiply \nyo by \nl
\advance \nxo by - \nl
\advance \nyo by - \nl
\multiply \ny by  \nl
\linethickness{0.3mm}
\multiply \nx by \nl
\multiput(\nxo,\nyo)(\nx,0){2}%
{\line(0,1){\ny}}
\nx=#3
\ny=#4
\multiply \nx by  \nl
\multiply \ny by  \nl
\multiput(\nxo,\nyo)(0,\ny){2}%
{\line(1,0){\nx}}
}
\def\putc#1#2{
\nxo=#1
\nyo=#2
\multiply \nxo by 4
\multiply \nyo by 4
\divide \nl by 2
\advance \nxo by - 2
\advance \nyo by -2
\thicklines
\put(\nxo,\nyo){\circle{1.5}}
\put(\nxo,\nyo){\circle{0.8}}
}
\def\putcc#1#2{
\nxo=#1
\nyo=#2
\nl=4
\multiply \nxo by \nl
\multiply \nyo by \nl
\divide \nl by 2
\advance \nxo by - \nl
\advance \nyo by - \nl
\thicklines
\put(\nxo,\nyo){\circle{2.0}}
\put(\nxo,\nyo){\circle{1.5}}
\put(\nxo,\nyo){\circle{0.8}}
}
\def\const#1#2{
\nxo=#1
\nyo=#2
\multiply \nxo by 4
\multiply \nyo by 4
\advance \nxo by - 2
\advance \nyo by - 2
\put(\nxo,\nyo){\circle*{2.7}}
}

\def\constx#1#2{
\setlength{\unitlength}{0.3mm}
\nxo=#1
\nyo=#2
\linethickness{0.8mm}
\multiply \nxo by 20
\multiply \nyo by 20
\advance \nxo by - 19%
\advance \nyo by - 18%
\put(\nxo,\nyo){\line(1,0){18}}%
\advance \nyo by 16
\put(\nxo,\nyo){\line(1,0){18}}
\advance \nyo by -17
\advance \nxo by 1
\put(\nxo,\nyo){\line(0,1){18}}
\advance \nxo by 16
\put(\nxo,\nyo){\line(0,1){18}}
\setlength{\unitlength}{1.5mm}
}
\long\def\symbolfootnote[#1]#2{\begingroup%
\def\thefootnote{\fnsymbol{footnote}}\footnote[#1]{#2}\endgroup} 

\numberwithin{equation}{section}

\begin{document}

\thispagestyle{empty}
\begin{flushright}
\end{flushright}
\baselineskip=16pt
\vspace{.5in}
{
\begin{center}
{\bf Partition Functions for Maxwell Theory on the Five-torus}

{\bf and for the Fivebrane on ${\bf S^1\times T^5}$}
\end{center}}
\vskip 1.1cm
\begin{center}
\centerline{ Louise Dolan\symbolfootnote[1] 
{E-mail:ldolan@physics.unc.edu}
and Yang Sun\symbolfootnote[2]{E-mail: sylmf@email.unc.edu}}
\bigskip
\centerline{\em Department of Physics}
\centerline{\em University of North Carolina, Chapel Hill, NC 27599} 
\bigskip
\vskip5pt

\bigskip
\bigskip
\bigskip
\bigskip
\end{center}

\abstract{\noindent 
We compute the partition function of five-dimensional abelian gauge
theory on a five-torus $T^5$ with a general flat metric using 
the Dirac method of quantizing with constraints. 
We compare this with the partition function of a single
fivebrane compactified on $S^1$ times $T^5$, which is obtained
from the six-torus calculation of Dolan and Nappi
[\href{http://xxx.lanl.gov/abs/hep-th/9806016}
{{\tt arXiv:hep-th/9806016}}]. The radius $R_1$ of the circle $S^1$ 
is set to the dimensionful gauge coupling constant $g^2_{5YM}= 4\pi^2 R_1$.
We find the two partition functions are equal only in the limit
where $R_1$ is small relative to $T^5$, a limit which removes the 
Kaluza-Klein modes from the 6d sum. This suggests the 6d $N=(2,0)$ tensor 
theory on a circle is an ultraviolet completion of the 5d gauge theory,
rather than an exact quantum equivalence. \Black
\bigskip}

\setlength{\parindent}{0pt}
\setlength{\parskip}{6pt}

\setstretch{1.05}
\vfill\eject
\vskip50pt
\section{\bf Introduction}

A quantum equivalence between the six-dimensional $N=(2,0)$ theory of multiple
fivebranes compactfied on a circle $S^1$
and five-dimensional maximally supersymmetric Yang Mills has been conjectured
by Douglas and Lambert {\it et al.}
in \cite{Douglas, Lambert}. In this paper we will study an abelian
version of the conjecture where the common five-manifold is a five-torus
$T^5$ with a general flat metric, and find an equivalence only in the weak
coupling limit. 

The physical degrees of freedom
of a single fivebrane are described by an $N=(2,0)$ tensor supermultiplet
which includes a chiral two-form field potential, so even a single
fivebrane has no fully covariant action. In order to investigate its 
quantum theory we were thus led in \cite{DN} to compute the partition
function instead, which we carried out on the six-torus $T^6$. 
We will use this calculation to investigate the partition function
of the self-dual three-form field strength restricted to $S^1\times T^5$
and compare it with the partition function of the five-dimensional
Maxwell theory on a twisted five-torus quantized via Dirac constraints
in radiation gauge. 

Because both the theory and the manifold are so simple, we do not
use localization techniques fruitful for non-abelian theories
and their partition functions on spheres \cite{WittenBeasley}-\cite{Tachikawa}. 

The five-dimensional Maxwell partition function on $T^5$
is defined\footnote{
Related work is 
\cite{BG} which appeared after an earlier version of this paper. See
also \cite{Gustavsson},\cite{Dijkgraaf}.}
as in string theory \cite{GSW}, 
\begin{align} Z^{5d, Maxwell}&\equiv
tr e^{-2\pi H^{5d} + i 2\pi \gamma^i P^{5d}_i} =
Z^{5d}_{\rm zero\;modes}\;\cdot Z^{5d}_{\rm osc},\cr
H^{5d}&= {R_6\over g^2_{5YM}}\int_0^{2\pi} d\theta^2d\theta^3d\theta^4
d\theta^5 \sqrt{g}\;
\big( {1\over 2 R^2_6} g^{ii'}\,F_{6i}F_{6i'}
+ {1\over 4} g^{ii'}g^{jj'} F_{ij}F_{i'j'} \big),\cr
&P^{5d}_i =  {1\over g^2_{5YM} R_6} 
\int_0^{2\pi} d\theta^2 d\theta^3 d\theta^4 d\theta^5 
\sqrt{g}\; 
g^{jj'}\,F_{6j'}F_{ij},
\label{wholepf}\end{align}
in terms of the gauge field strength $F_{\tilde m\tilde n} 
(\theta^2,\theta^3,\theta^4,\theta^5,\theta^6),$
and constant metric $g^{ij}, R_6, \gamma^i$. 
The partition function of the abelian chiral two-form on a space circle times
the five-torus is
\begin{align}
Z^{6d, chiral} &=tr \,e^{-2\pi R_6{\cal H} + i2\pi\gamma^i{\cal P}_i}
= Z^{6d}_{\rm zero \; modes}  \cdot Z^{6d}_{\rm osc},\cr
{\cal H} &= {1\over 12}\int_0^{2\pi} d\theta^1 \ldots d\theta^5 
{\sqrt G_5}{G_5}^{ll'}
{G_5}^{mm'}{G_5}^{nn'} H_{lmn}(\vec\theta,\theta^6)\,
H_{l'm'n'}(\vec\theta,\theta^6),\cr
{\cal P}_i&= -{1\over 24}{\int_0}^{2\pi}
d\theta^1...d\theta^5 \epsilon^{rsumn} H_{umn}(\vec\theta,\theta^6)\,
H_{irs}(\vec\theta,\theta^6)
\label{6dpf}\end{align}
where $\theta^1$ is the direction of the circle $S^1$.
The time direction $\theta^6$ we will use for quantization 
is common to both theories, and the
angles between the circle and the five-torus denoted by $\alpha,\beta^i$ in 
\cite{DN} have been set to zero. The final 
results are given in (\ref{wpfyetagain}), (\ref{wpfmore}). 

We use (\ref{wholepf},\ref{6dpf}) 
to compute both the zero mode and 
oscillator contributions, and find an exact equivalence 
between the zero mode contributions,
\begin{align}
Z^{6d}_{\rm zero \; modes} 
= Z^{5d}_{\rm zero \; modes}.
\label{zerocomp}\end{align}
Not surprisingly, we find the oscillator traces differ by the absence in 
$Z^{5d}_{\rm osc}$ of the 
Kaluza-Klein modes generated in $Z^{6d}_{\rm osc}$
from compactification on the circle $S^1$. 

The Kaluza-Klein modes have been associated with instantons in 
the five-dimensional non-abelian gauge theory in \cite{Douglas,Lambert,
KimyeongLee, CollieTong}, with additional comments given for the abelian limit.
It would be interesting to find a systematic way to incorporate
these modes in a generalized five-dimensional partition function along the 
lines of a character, in order to match the partition functions
exactly, but we have not done that here. Rather
our explicit expressions show an equivalence between the oscillator traces
of the two theories only in the limit where the compactification
radius $R_1$ of the circle is small compared to the five-torus $T^5$.

Other approaches to $N=(2,0)$ theories
formulate fields for non-abelian chiral two-forms \cite{Singh}-\cite{Lee}
which would be useful if 
the non-abelian six-dimensional theory has
a classical description and if the quantum theory
can be described in terms of fields. On the other hand the 
partition functions on various manifolds \cite{Nekrasov}-\cite{Vandoren}
can demonstrate aspects of 
the six-dimensional finite quantum conformal theory presumed 
responsible for features of four-dimensional gauge theory \cite{Witten}.

In section 2, the contribution of the zero modes to the partition function
for the
chiral theory on a circle times a five-torus is computed as a sum over the ten
integer eigenvalues, and its relation to that of the gauge theory 
is shown via a fiber bundle approach. 
In section 3, the abelian gauge theory is quantized on 
a five-torus using Dirac constraints, and the Hamiltonian and momenta
are computed in terms of the oscillator modes.
In section 4, we construct the oscillator trace contribution to the
partition function for the gauge theory 
and compare it with that of the chiral two-form. Section 5 contains
discussion and conclusions. Appendix A presents details of the Dirac
quantization and Appendix B verifies the Hamilton equations of motion.
Appendix C regularizes the vacuum energy.
Appendix D proves the $SL(5,{\cal Z})$ invariance of both partition functions.
\section{\bf Zero Modes}
\label{zeromodes}

The $N=(2,0)$ 6d world volume theory of the fivebrane
contains five scalars, two four-spinors
and a chiral two-form $B_{MN}$, 
which has a self-dual three-form field strength
$H_{LMN}= \partial_LB_{MN} + \partial_MB_{NL} + \partial_NB_{LM}$
with $1\le L,M,N \le 6$,
\begin{align}
H_{LMN}(\vec\theta,\theta^6)= {1\over{6\sqrt{-G}}}G_{LL'}G_{MM'}G_{NN'}
\epsilon^{L'M'N'RST}H_{RST}(\vec\theta,\theta^6).
\label{sd3form}\end{align}
(\ref{sd3form}) gives 
$H_{LMN}(\vec\theta,\theta^6)= {{\textstyle i}
\over{6\sqrt{|G|}}}G_{LL'}G_{MM'}G_{NN'}
\epsilon^{L'M'N'RST}H_{RST}(\vec\theta,\theta^6)$ for a 
Euclidean signature metric.
In the absence of a covariant Lagrangian, the partition function
of the chiral field is defined via a trace over the Hamiltonian \cite{DN} 
as is familiar from string calculations. We display this expression in 
(\ref{6dpf}) where the metric has been restricted to describe
the line element for $S^1\times T^5$,
\begin{align}
ds^2 &= {R_1}^2(d\theta^1)^2 + {R_6}^2(d\theta^6)^2
+ \sum_{i,j=2...5}g_{ij}(d\theta^i -\gamma^id\theta^6)
(d\theta^j - \gamma^jd\theta^6) 
\label{lineelem}\end{align}
with $0\le\theta^I\le 2\pi$, $1\le I\le 6$. 
The parameters $R_1$ and $R_6$ are the radii for
directions 1 and 6,
$g_{ij}$ is a 4d metric, and $\gamma^j$ are the angles between
between 6 and $j$. So from (\ref{lineelem}),
\begin{align}
&{G}_{ij}= g_{ij}\,;
\;G_{11}={R_1}^2;\quad
G_{i1}= 0\,;\quad
G_{66}= {R_6}^2 + g_{ij}\gamma^i\gamma^j;
\quad G_{i6}=-g_{ij}\gamma^j\,;
\quad G_{16}=0;
\label{sixmetric}\end{align}
and the inverse metric is\begin{align}
&G^{ij}= g^{ij}+{\gamma^i\gamma^j\over R_6^2};
\quad G^{11}={1\over R^2_1};\quad
G^{1i}=0;\quad
G^{66}={1\over R_6^2}; 
\quad G^{i6}={\gamma^i\over R_6^2};\quad G^{16}=0.
\label{sixmetricinverse}\end{align}
We want to keep the time direction $\theta^6$ common to both theories,
so in the 5d expressions (\ref{wholepf}) the indices are on 
$2\le\tilde m,\tilde n\le 6$;
whereas the Hamiltonian and momenta in (\ref{6dpf}) 
sum on $1\le m,n\le 5$. 
The common space index is labeled $2\le i,j\le 5$.
To this end, for the metric
$G_{MN}$ in (\ref{sixmetric}) we introduce
the 5-dimensional inverse (in directions 1,2,3,4,5) 
\begin{align}{G_5}^{ij}= g^{ij};\qquad
{G_5}^{i1}=0;\qquad
{G_5}^{11}= {1\over R_1^2};\end{align}
and the 5-dimensional inverse (in directions 2,3,4,5,6) 
for the five-torus $T^5$, \begin{align}
{\widetilde G_5}^{ij}=& g^{ij} + {\gamma^i\gamma^j\over R_6^2}\,;\qquad
{\widetilde G_5}^{i6}={\gamma^i\over R_6^2}\,;\qquad
{\widetilde G_5}^{66}= {1\over R_6^2}.
\label{tildeg5}\end{align}
The determinants of the metrics are related simply by
$\sqrt{G} = R_6\sqrt{G_5} = R_1\sqrt{\widetilde G_5} = R_6 R_1\sqrt{g}$\,,
and $\epsilon_{23456} \equiv \widetilde G_5 \,
\epsilon^{23456} =\widetilde G_5$,
with corresponding epsilon tensors related by $G$, $G_5$, $g$.

To compute $Z_{\rm zero\, modes}^{6d}$ we neglect the integrations
in (\ref{6dpf}) and get
\begin{align}
-2\pi R_6{\cal H}
&=-{\pi \over 6}  R_6R_1{\sqrt g}g^{ii'}g^{jj'}g^{kk'}
H_{ijk}H_{i'j'k'}
-{\pi\over 4} {R_6\over R_1}{\sqrt g}(g^{jj'}g^{kk'}
- g^{jk'}g^{kj'}) H_{1jk}H_{1j'k'},\cr
i2\pi\gamma^i \P_i &=
-{i\pi \over 2} \gamma^i \epsilon^{jkj'k'} H_{1jk} H_{ij'k'} 
= {i\pi\over 3} \gamma^i \epsilon^{jj'kk'}  H_{j'kk'}H_{1ij},
\label{pmo}\end{align}
\vskip-15pt
where the zero modes of the four fields $H_{ijk}$ are labeled by the integers
$n_7,\ldots , n_{10}$. The six fields $H_{1jk}$ have zero mode eigenvalues
$H_{123}=n_1$, $H_{124}=n_2$, $H_{125}=n_3$, $H_{134}=n_4$, $H_{135}=n_5$,
$H_{145}=n_6$, and the trace on the zero mode operators in  (\ref{6dpf}) is

\begin{align}
Z_{\rm zero\, modes}^{6d}&=
\sum_{n_1,\ldots,n_6} 
\exp\{-{\pi\over 4} {R_6\over R_1}{\sqrt g}
(g^{jj'}g^{kk'} - g^{jk'}g^{kj'}) H_{1jk}H_{1j'k'}\}
\cr
&\hskip10pt\cdot
\sum_{n_7,\ldots,n_{10}}
\exp\{-{\pi \over 6}  R_6R_1{\sqrt g}g^{ii'}g^{jj'}g^{kk'}
H_{ijk}H_{i'j'k'}
-{i\pi \over 2}\gamma^i \epsilon^{jkj'k'} H_{1jk} H_{ij'k'}\}. \cr
\label{chzm}\end{align}
The same sum is obtained from the 5d Maxwell theory
(\ref{wholepf}) where the gauge coupling is identified\footnote{
See for example arXiv:1012.2882, p5 \cite{Lambert}.} with the
radius of the circle $g^2_{5YM} = 4\pi^2 R_1$, as follows. 
The zero modes of the gauge theory are eigenvalues of operator-valued 
fields that satisfy Maxwell equations with no sources. Even classically
these solutions have constant
$F_{ij}$ which lead to non-zero flux through closed two-surfaces that
are not a boundary of a three-dimensional submanifold in $T^5$.
Working in $A_6=0$ gauge,
if we consider the $U(1)$ gauge field $A_i$
at any time $\theta^6$ as a connection on a principal U(1) bundle
with base manifold $T^4$, then the curvature 
$F_{ij}=\partial_iA_j-\partial_jA_i$\break for $2\le i,j\le 5$ must have
integer flux \cite{Wittentwo,Verlinde}, in the sense that
\begin{align}
n_I = {1\over 2\pi}\int_{\Sigma_2^I} F\equiv  {1\over 2\pi}\int_{\Sigma_2^I}
{1\over 2} F_{ij} \, d\theta^i\wedge d\theta^j,\qquad n_I\in \Z,
\; \hbox{for each} \; 1\le I\le 6.\label{coho}
\end{align} 
In $T^4$, the six representative two-cycles $\Sigma_2^I$ are each a 2-torus
constructed by the six ways of combining the four $S^1$ of $T^4$ two at a time,
given by the cohomology class, $\dim H_2(T^4) = 6$. Relabeling
$n_I$ as $n_{i,j}$ and $\Sigma_2^I$ as $\Sigma_2^{i,j}$, $2\le i< j\le 5$, we have 
$\int_{\Sigma_2^{g,h}} d\theta^i\wedge d\theta^j 
= (2\pi)^2 (\delta^i_g\delta^j_h
- \delta^i_h\delta^j_g)$. So ({\ref{coho}) is
\begin{align}
F_{ij} = {n_{i,j}\over 2\pi}, \qquad n_{i,j}\in\Z \; \hbox{for $i<j$}.  
\label{Fint}\end{align}  
Furthermore we show how the zero mode
eigenvalues of $F_{6i}$ are found\footnote{This point of view is discussed 
in \cite{Henningson}. See also \cite{BG}.} 
from those of the conjugate momentum $\Pi^i$.
In section 3 we derive the form of $H^{5d}$ and $P^{5d}_i$ given in
(\ref{wholepf}) from
a canonical quantization using a Lorentzian signature metric.
In (\ref{conjmomL}) the conjugate momentum is defined as 
\begin{align}
\Pi^i &= {\sqrt{g}\over 4\pi^2 R_1R_6} g^{ii'}F_{6i'}.
\label{fmom}\end{align} 
From the commutation relations (\ref{thcr}) we can compute its commutator
with the holonomy $\int_{\Sigma_1^k} A \equiv
\int_{\Sigma_1^k} A_i(\vec\theta,\theta^6) d\theta^i$
where $\Sigma_1^k$ are the four representative one-cycle circles in $T^4$,
\begin{align}
\left[\int_{\Sigma_1^k} A_i(\vec\theta,\theta^6) d\theta^i,
\int {d^4\theta'\over 2\pi}  \Pi^j(\vec\theta',\theta^6)\right]
&= {i\over 2\pi} \int_{\Sigma_1^k} d\theta^j = i \,\delta_k^j.
\end{align}
Hence an eigenstate $\psi$ of the the zero mode operator
${1\over 2\pi} \int d^4\theta' 
\Pi^k(\vec\theta',\theta^6)$ with eigenvalue $\lambda$ is
\begin{align}
\psi = e^{i \lambda\int_{\Sigma_1^k}A} \, |0\rangle,\qquad
\Big({1\over 2\pi}\int d^4\theta' \Pi^k(\vec\theta',\theta^6)\Big) \;
e^{i \lambda\int_{\Sigma_1^k}A} \, |0\rangle
= \lambda \; \, e^{i \lambda\int_{\Sigma_1^k}A} \, |0\rangle.
\nonumber\end{align}
Since the holonomy is defined mod $2\pi$, thus allowing $A$ to vary by gauges
when crossing neighborhoods, but ensuring 
$e^{i \int_{\Sigma_1^k}A}$ to be a single valued element of 
the structure group $U(1)$, then the states
\begin{align}
e^{i \lambda \int_{\Sigma_1^k}A} \, |0\rangle \qquad \hbox{and}\qquad
e^{i \lambda \big( 2\pi + \int_{\Sigma_1^k}A\big)} \, |0\rangle
\end{align}
must be equivalent, so the eigenvalue $\lambda$ of
the operator ${1\over 2\pi} \int d^4\theta' \Pi^k(\vec\theta',\theta^6)$ 
must have integer values $n^{(k)}$,
\begin{align}
\Pi^k(\vec\theta',\theta^6)
= {n^{(k)}\over (2\pi)^3},\qquad n^{(k)}\in \Z^4.
\label{piint}\end{align}
In this normalization of the zero mode eigenvalues for the gauge theory, we
are taking the $d\theta^i$ space integrations into account.
So (\ref{wholepf}) gives 
\begin{align}
&- 2\pi H^{5d} + i 2\pi \gamma^i P_i^{5d}\cr
&= \Big( - {\pi\sqrt{g}\over R_1R_6} g^{ii'} F_{6i} F_{6i'} 
-{\pi R_6\over 2 R_1}\sqrt{g} g^{ii'}g^{jj'} F_{ij}
F_{i'j'} + 2\pi i \gamma^i {\sqrt{g}\over R_1R_6} g^{jj'}
F_{6j'}F_{ij}\Big) \, (2\pi)^2.\cr
\label{maxwell5}
\end{align}
We can use the identity 
\begin{align}
-{1\over 4} \epsilon^{jkj'k'} H_{1jk}H_{ij'k'} &=
{1\over 6} \epsilon^{jj'kk'}  H_{j'kk'}H_{1ij},
\nonumber\end{align}
to rewrite the last term in (\ref{chzm}) as
\begin{align}
-{i\pi \over 2} \gamma^i \epsilon^{jkj'k'} H_{1jk} H_{ij'k'}
&=  {i\pi \over 3} \gamma^i \epsilon^{jj'kk'} H_{j'kk'}H_{1ij},
\nonumber\end{align}
which is equal to the last term in (\ref{maxwell5}) if we identify
\begin{align}
{1\over 6} \epsilon^{jj'kk'} H_{j'kk'} =  {2\pi \sqrt{g}\over R_1R_6}
g^{jj'} F_{6j'},\qquad H_{1ij} = 2\pi F_{ij}.
\label{newid}\end{align}
Then, from (\ref{newid}) we have
that the first term in (\ref{maxwell5}) becomes\footnote{
\begin{align}
&g_{jg} \epsilon^{jj'kk'} \epsilon^{gg'hh'}
= g( g^{j'g'}g^{kh}g^{k'h'} - g^{j'g'}g^{k'h}g^{kh'}
- g^{k'g'}g^{kh}g^{j'h'} + g^{k'g'}g^{j'h}g^{kh'}
- g^{k g'}g^{j'h}g^{k'h'} + g^{kg'}g^{k'h}g^{j'h'}),\cr
&\epsilon^{2345} =1 \;\hbox{and}\; \epsilon_{2345} = g \epsilon^{2345} = g.
\nonumber\end{align}}

\begin{align}
-{4\pi^3\sqrt{g}\over R_1R_6} g^{ii'} F_{6i}F_{6i'}
&= -{\pi\over 6}\sqrt{g}R_1R_6 \;g^{j'g'}g^{gh}g^{g'h'} H_{j'kk'}H_{g'hh'}.
\nonumber\end{align}
Thus with the identifications in (\ref{newid}),
the 5d Maxwell expression in (\ref{maxwell5}) is equal to 
the 6d chiral exponent in (\ref{chzm}),
\begin{align}
-2\pi H^{5d} + i2\pi\gamma^i P^{5d}_i &= \Big(
-{\pi \sqrt{g}\over R_1 R_6} g^{ii'}\,F_{6i}F_{6i'}
-{\pi R_6\sqrt{g}\over 2 R_1} g^{ii'}g^{jj'} F_{ij}F_{i'j'} 
+ {i 2\pi \sqrt{g}\over R_1 R_6} \gamma^i g^{jj'}\,F_{6j'}F_{ij}\Big)
(2\pi)^2 \cr
= -t{\cal H} + i2\pi\gamma^i\P_i
&= -{\pi \over 6}  R_6R_1{\sqrt  g}g^{ii'}g^{jj'}g^{kk'}
H_{ijk}H_{i'j'k'}
-{\pi\over 4}{R_6\over R_1}\sqrt{g} (g^{jj'}g^{kk'}-g^{jk'}g^{j'k})
H_{1jk}H_{1j'k'}\cr
&\hskip15pt -{i\pi \over 2} \gamma^i \epsilon^{jkj'k'} H_{1jk} H_{ij'k'}.
\nonumber\end{align}
We now discuss the sum over integers in (\ref{chzm}). From 
(\ref{newid}),
if $H_{1jk}$ are integers, then $2\pi \, F_{ij}$ are integers.
If $H_{ijk}$ are integers, then ${1\over 6}\epsilon^{jj'kk'} H_{j'kk'}$
are also integers. This implies, again from  (\ref{newid}), that
${2\pi\,\sqrt{g}\over R_1R_6} g^{jj'} F_{6j'}$ should be integers, which we
justify in (\ref{Fint}) and (\ref{piint}) with (\ref{fmom}). 
Thus the Maxwell zero mode trace can be written as
\begin{align}
Z^{5d}_{\rm zero\;modes}&= \sum_{n_1\ldots n_6}
\,\exp\{-{2\pi^3}{R_6\sqrt{g}\over R_1} g^{ii'}g^{jj'}F_{ij}F_{i'j'}
\} \cr
&\hskip40pt\cdot
\sum_{n^7\ldots n^{10}} \exp\{-{4\pi^3\sqrt{g}\over R_1R_6} g^{ii'}
F_{6i} F_{6i'} 
+ {i (2\pi)^3 \sqrt{g}\over R_1 R_6} \gamma^i g^{jj'}\,F_{6j'}F_{ij} \}
\label{pfi}\end{align}
where the integer eigenvalues are 
$n_1= 2\pi F_{23}, n_2=2\pi F_{24}, n_3= 2\pi F_{25}, n_4=2\pi F_{34},
\hfill\break
n_5=2\pi F_{35}, n_6=2\pi F_{45}$;
$(n^7,n^8,n^9,n^{10}) \equiv (n^{(2)},n^{(3)},n^{(4)},n^{(5)})$,\hfill\break
for
$n^{(k)}\equiv {2\pi\sqrt{g}\over R_1R_6} g^{ki'}F_{6i'}\in\Z^4.$
So we have 
proved the relation (\ref{zerocomp})
\begin{align}
Z^{6d}_{\rm zero \; modes} = Z^{5d}_{\rm zero \; modes}
\label{zerocompa}\end{align}
and the explicit expression is given by (\ref{chzm}) or (\ref{pfi}).

\section{\bf Dirac Quantization of Maxwell Theory on a Five-torus}
\label{diracquant}

To evaluate the oscillator contribution to the partition function 
in (\ref{wholepf}), we will
first quantize the abelian gauge theory on the five-torus with
a general metric. The equation of motion is
$\partial^{\tilde m} F_{\tilde m\tilde n} = 0.$ For
$ F_{\tilde m\tilde n} = \partial_{\tilde m} A_{\tilde n}
- \partial_{\tilde n} A_{\tilde m}$, a solution 
is given by a solution to
\begin{align}
\partial^{\tilde n}\partial_{\tilde n} A_{\tilde m} = 0,\qquad
\partial^{\tilde m}A_{\tilde m} = 0.
\label{eomagt}\end{align}
These have a plane wave solution
$A_{\tilde m}(\vec\theta,\theta^6) = f_{\tilde m}(k) e^{ik\cdot\theta}
+ (f_{\tilde m}(k) e^{ik\cdot\theta})^\ast$\;
when
\begin{align}
\widetilde G_L^{\tilde m \tilde n} k_{\tilde m}k_{\tilde n} = 0,
\qquad k^{\tilde m} f_{\tilde m} =0.
\label{5dg}\end{align}
In order for the operator formalism (\ref{wholepf}) to reproduce
a path integral quantization with spacetime metric (\ref{tildeg5}), we 
must canonically quantize $H^{5d}$ and $P^{5d}_i$ via a metric that has 
zero angles with the time
direction, {\it i.e.} $\gamma^i=0$, and insert $\gamma^i$ in the partition
function merely as the coefficient of  $P^{5d}_i$ \cite{GSW}.
Furthermore a Lorentzian signature metric is needed for
quantum mechanics, so we modify the metric on the five-torus
(\ref{sixmetric}), (\ref{tildeg5}) to be
\begin{align}&{\widetilde G}_{L\,ij}= g_{ij}\,;\;
\widetilde G_{L\,66}= -{R_6}^2;
\; \widetilde G_{L\,i6}=0\,;\quad
\widetilde G_L^{ij}= g^{ij};\;
\widetilde G_L^{66}=-{1\over R_6^2};
\;\widetilde G_L^{i6}=0,\quad \widetilde G_L = \det \widetilde 
G_{L\,\tilde m\tilde n}.
\label{Lfivemetric}\end{align}
Solving for $k_6$ from
(\ref{5dg}) we find
\begin{align}
k_6= \pm {\sqrt{-\widetilde G_L^{66}}\over \widetilde G_L^{66}}\,
|k|, \label{k6}\end{align}
where $2\le i,j\le 5,$ and $|k| \equiv \sqrt{g^{ij} k_ik_j}.$
Use the gauge invariance $f_{\tilde m}\rightarrow f'_{\tilde m}
= f_{\tilde m} + k_{\tilde m} \lambda$ to fix $f'_6=0,$
which is the gauge choice $$A_6=0.$$
This reduces the number of components of $A_{\tilde m}$ from 5 to 4.
To satisfy (\ref{5dg}), we can use the $\partial^{\tilde m} F_{\tilde m 6}
= -\partial_6\partial^i A_i=0$
component of the equation of motion to eliminate $f_5$ in terms of the three
$f_2,f_3,f_4$,
\begin{align}
f_5= -{1\over p^5}(p^2f_2+p^3f_3+p^4f_4),
\nonumber\end{align}
leaving just three independent polarization vectors
corresponding to the physical degrees of freedom of the 5d one-form
with Spin(3) content 3.
From the Lorentzian Lagrangian
\begin{align}
\cL= -{1\over 4}{\sqrt{-\widetilde G_L}\over g^2_{5YM}} 
\widetilde G_L^{\tilde m\tilde {m'}} 
\widetilde G_L^{\tilde n\tilde {n'}} F_{\tilde m\tilde n}
F_{\tilde m'\tilde n'}
= {R_6\sqrt{g}\over 4\pi^2R_1} \Big(-{1\over 4 }g^{ii'}g^{jj'} F_{ij} F_{i'j'} 
- {1\over 2} \widetilde G_L^{66}
g^{jj'} F_{6j} F_{6j'}\Big),\label{5dLor}
\end{align}
the energy-momentum tensor
\begin{align}
\T^{\,m}_{\hskip8pt n} &=  {\delta\cL\over \delta \partial_m A_p}
\partial_n A_p - \delta^m_{\hskip4pt_n}\,\cL 
\end{align} leads to the Hamiltonian and momenta operators
\begin{align}
H_c &\equiv \int d^4\theta \, \T^{\,6}_{\hskip8pt6} = \int d^4\theta\,
\Big( {R_6\sqrt{g}\over  4\pi^2 R_1}\big(
-{1\over 2} \widetilde G_L^{66}g^{ii'}\,F_{6i}F_{6i'}
+ {1\over 4} g^{ii'}g^{jj'} F_{ij}F_{i'j'}
- F^{6i}\partial_i A_6\big) + \Pi^6 \partial_6 A_6 \Big),\label{firstHC}\\
P_i&\equiv\int d^4\theta \,\T^{6}_{\hskip8pt i} =
\int d^4\theta \Big( {R_6\sqrt{g}\over  4\pi^2 R_1}
\big (-\widetilde G_L^{66}g^{jj'} \,F_{6j'}F_{ij} - F^{6j}\partial_j A_i \big)
+ \Pi^6 \partial_i A_6 \Big) \label{firstP},\cr
\end{align}
where the conjugate momentum is
\begin{align}
\Pi^i = {\delta\cL\over \delta \partial_6 A_i} = -{R_6\sqrt{g}\over 4\pi^2 R_1} F^{6i} = - {R_6\sqrt{g}\over  4\pi^2 R_1} \widetilde G_L^{66} g^{ii'} F_{6i'},
\qquad \Pi^6 =  {\delta\cL\over \delta \partial_6 A_6} = 0.
\label{conjmomL}
\end{align}
We quantize the Maxwell field on the five-torus with the metric 
(\ref{Lfivemetric}) in radiation gauge 
using Dirac constraints \cite{Dirac, Das}. 
The theory has a primary constraint $\Pi^6(\vec \theta, \theta^6) \approx 0.$
We can express the Hamiltonian (\ref{firstHC}) in terms of the conjugate
momentum as 
\begin{align}
H_{can}&= \int d^4\theta \Big( -{  2\pi^2 R_1 \over R_6\sqrt{g} 
\widetilde G_L^{66}}\, g_{ii'}\Pi^i\,\Pi^{i'} +
{R_6\sqrt{g}\over  16\pi^2 R_1} g^{ii'}g^{jj'} F_{ij} F_{i'j'}
- \partial_i\Pi^i \,A_6  \Big),
\label{Hcan}\end{align}
where the last term has been integrated by parts.
The primary Hamiltonian is defined by
\begin{align}
H_p&= \int d^4\theta \Big( -{ 2\pi^2 R_1 \over R_6\sqrt{g} 
\widetilde G_L^{66}}\, g_{ii'}\Pi^i\,\Pi^{i'} +
{R_6\sqrt{g}\over 16\pi^2 R_1} g^{ii'}g^{jj'} F_{ij} F_{i'j'}
- \partial_i\Pi^i \,A_6   + \lambda_1\Pi^6\Big),
\label{Hp}\end{align}
with $\lambda_1$ as a Lagrange multiplier. 
In Appendix A, we use the Dirac method of quantizing with constraints 
for the radiation gauge conditions $A_6\approx 0$,
$\partial^i A_i\approx 0$, and find the equal time commutation relations
(\ref{DirCom}), (\ref{othcom}):
\begin{align}
&[\Pi^j(\vec\theta, \theta^6), A_i(\vec\theta',\theta^6)] =
- i \Big( \delta^j_i - g^{jj'}
(\partial_i{1\over g^{kk'}\partial_k\partial_{k'}}
\partial_{j'})\Big) \;\delta^4 (\theta-\theta'),\cr
& [ A_i(\vec\theta,\theta^6), A_j(\vec\theta',\theta^6)]=0,
\qquad [\Pi^i(\vec\theta,\theta^6),\Pi^j(\vec\theta',\theta^6)]=0.
\label{thcr}\end{align}
Appendix B shows the Hamilitonian (\ref{Hp})  
to give the correct equations of motion.

In $A_6=0$ gauge,
the free quantum vector field on the torus is expanded as
\begin{align}
A_i(\vec\theta,\theta^6) = \;{\rm zero\, modes}\; + \sum_{\vec k\ne0, \vec k
\in {\cal Z}_4}
(f_i^{\kappa} a_{\vec k}^{\kappa} e^{ik\cdot \theta}
+ f_i^{\kappa\ast} a_{\vec k}^{\kappa\dagger} e^{-ik\cdot \theta}),
\nonumber\end{align}
where $1\le \kappa\le 3$, $2\le i\le 5$ and $k_6$ defined in (\ref{k6}).
The sum is on the dual lattice $\vec k = k_i\in {\cal Z}_4\ne\vec 0.$
Having computed the zero mode contribution in (\ref{pfi}), 
here we consider\footnote{In this mode expansion, we shall pick the plus sign
in (\ref{k6}) for the root $k_6$ which solves (\ref{5dg}).}
\begin{align}
A_i(\vec\theta,\theta^6) = \sum_{\vec k\ne 0}
(a_{\vec k\, i} e^{ik\cdot \theta}
+ a_{\vec k\, i}^{\dagger} e^{-ik\cdot \theta}),
\label{vecfield}\end{align}
with polarizations absorbed in 
\begin{align}
a_{\vec k\,i}= f_i^\kappa a_{\vec k}^\kappa.\label{polarize}
\end{align}
From (\ref{thcr}) the commutator in terms of the oscillators is 
\begin{align}
\int {d^4\theta d^4\theta'\over (2\pi)^8}
e^{- i k_i\theta^i} e^{- i {k'}_i{\theta'}^i}
[ A_i(\vec\theta, 0), A_j(\vec\theta', 0)]
= [(a_{\vec k\, i} + a_{-\vec k\, i}^\dagger),
(a_{\vec k'\, j} + a_{-\vec k'\, j}^\dagger)] = 0.
\label{Acom}\end{align}
The conjugate momentum $\Pi^j(\vec\theta,\theta^6)$ in (\ref{conjmomL})
is expressed in terms of $a_{\vec k\,i}, a_{-\vec k\,i}^\dagger$ by
\begin{align}
\Pi^j(\vec\theta,\theta^6) 
&= -i{R_6\sqrt{g}\over 4\pi^2 R_1} \widetilde G_L^{66} g^{jj'}
\sum_{\vec k} k_6 
\, (a_{\vec k\, j'} e^{ik\cdot\theta}
- a_{\vec k\, j'}^\dagger e^{-ik\cdot\theta}).
\label{pie}\end{align}
Then taking the Fourier transform of $\Pi^j(\vec\theta,\theta^6)$
at $\theta^6=0$, we have
\begin{align}
\int {d^4\theta\over (2\pi)^4} e^{-ik_i\theta^i} \Pi^j(\vec\theta, 0)
&= - i{R_6\sqrt{g}\over 4\pi^2 R_1} \widetilde G_L^{66} g^{jj'} 
k_6 \, (a_{\vec k\, j'} - a_{-\vec k\, j'}^\dagger).
\label{pia}\end{align}
From (\ref{pia}) and the commutators (\ref{thcr}) and (\ref{Acom}), we find
\begin{align}
&\int {d^4\theta d^4\theta'\over (2\pi)^8} e^{-ik_i\theta^i}
e^{-i{k'}_i{\theta'}^i} [\Pi^j(\vec\theta, 0),
A_i(\vec\theta', 0)] \cr&= -i(\delta^j_i - {g^{jj'}
k_ik_{j'}\over g^{kk'}k_kk_{k'}}) \delta_{\vec k,-\vec k'}\;
{1\over (2\pi)^4}
= - i{R_6\sqrt{g} \over 4\pi^2 R_1} \widetilde G_L^{66} g^{jj'} 
k_6 \, 
[(a_{\vec k\, j'} - a_{-\vec k\, j'}^\dagger),
(a_{\vec k'\, i} + a_{-\vec k'\, i}^\dagger)].\cr
\label{piatwo}\end{align}
To reach the oscillator commutator (\ref{aacom}), we define
\begin{align}
& A_{\vec k\, i} \equiv a_{\vec k\,i} + a^\dagger_{-\vec k\,i}
= A^\dagger_{-\vec k\, i},\qquad
 E_{\vec k\, i} \equiv a_{\vec k\,i} - a^\dagger_{-\vec k\,i}
= -E^\dagger_{-\vec k\, i},
\label{EA}\\
&a_{\vec k\,i} = {1\over 2} (A_{\vec k\,i} + E_{\vec k\,i}),
\qquad
a^\dagger_{\vec k\,i} = {1\over 2} (A^\dagger_{\vec k\,i} +
E^\dagger_{\vec k\,i}) = {1\over 2} (A_{-\vec k\,i} -
E_{-\vec k\,i}).
\label{aa}\end{align}
Now inverting (\ref{piatwo}) we have
\vskip-30pt \begin{align}
[E_{\vec k\, j}, A_{\vec k'\, i}] =
{R_1\over R_6\sqrt{g}\widetilde G_L^{66} k_6} {1\over (2\pi)^2}
\big( g_{ji} - {k_jk_i\over g^{kk'}k_kk_{k'}} \big) \delta_{\vec k,-\vec k'},
\label{piathree}\end{align}
and from  (\ref{pia}) and the relations (\ref{thcr}) and (\ref{Acom}),
\begin{align}
[A_{\vec k\, i}, A_{\vec k'\, j}] = 0,\qquad
[E_{\vec k\, i}, E_{\vec k'\, j}] &= 0.
\label{aaee}\end{align}
Using (\ref{aa}),
\begin{align}
[a_{\vec k\,i}, a^\dagger_{\vec k'\,j}] &=
{1\over 4}\Big( [A_{\vec k\,i}, A_{-\vec k'\,j}] -
[E_{\vec k\,i}, E_{-\vec k'\,j}] - [A_{\vec k\,i}, E_{-\vec k'\,j}]
+ [E_{\vec k\,i}, A_{-\vec k'\,j}]\Big),
\end{align}
together with (\ref{piathree}), (\ref{aaee})
we find the oscillator commutation relations
\begin{align}
[a_{\vec k\,i}, a^\dagger_{\vec k'\,j}] 
&= {R_1\over R_6\sqrt{g}\widetilde G_L^{66} k_6} {1\over 2 (2\pi)^2}
\big(  g_{ij} -  {k_ik_j\over g^{kk'}k_kk_{k'}} \big)
\delta_{\vec k,\vec k'},\cr
[a_{\vec k\,i}, a_{\vec k'\,j}]&=0,\qquad
[a^\dagger_{\vec k\,i}, a^\dagger_{\vec k'\,j}]=0.\label{aacom}
\end{align}
In the gauge  $\partial^i A_i(\vec\theta,\theta^6) =0$, then
$k^ia_{\vec k\,i} = g^{ij}k_j a_{\vec k\,i} = 0, \;k^ia^\dagger_{\vec k\,i}
= g^{ij} k_j a^\dagger_{\vec k\, i} = 0$ as in ({\ref{5dg}),
and these are consistent
with the commutator (\ref{aacom}).
We will use this commutator 
to proceed with the evaluation of the 
Hamiltonian and momenta in (\ref{firstHC},\ref{firstP}).
In $A_6=0$ gauge, 
\begin{align}
H_c &= \int d^4\theta\, {R_6\sqrt{g}\over 4\pi^2 R_1}\Big(
-{1\over 2} \widetilde G_L^{66}g^{ii'}\,\partial_6 A_i \partial_6 A_{i'}
+ {1\over 4} g^{ii'}g^{jj'} F_{ij}F_{i'j'}\Big),
\label{Hq}\end{align}
which is the Hamiltonian $H^{5d}$ in (\ref{wholepf}).
In (\ref{firstP}) after integrating by parts, we also set the 
second constraint described in Appendix A \; $\partial_i\Pi^i= 0$,
to find 
\begin{align}
P_i =  {1\over 4\pi^2 R_1 R_6}
\int_0^{2\pi} d\theta^2 d\theta^3 d\theta^4 d\theta^5
\sqrt{g}\;
g^{jj'}\,F_{6j'}F_{ij},
\label{Pq}\end{align}
which is the momenta $P^{5d}_i$ in (\ref{wholepf}).

From (\ref{Hq}), in terms of the normal mode expansion (\ref{vecfield}),
\begin{align}
H_c &= 
(2\pi)^2{R_6\sqrt{g}\over R_1}\sum_{\vec k\in\Z^4 \ne \vec 0}
\big( {1\over 2} \widetilde G_L^{66}g^{ii'}
k_6 k_6 + {1\over 2} (g^{ii'}g^{jj'}- g^{ij'}g^{ji'}) k_jk_{j'} \big)
(a_{\vec k\,i} a_{-\vec k\,i'}
e^{2 i k_6\theta^6}
+ a^\dagger_{\vec k\,i}a^\dagger_{-\vec k\,i'}
e^{- 2 i k_6\theta^6})\cr
&\hskip10pt + (2\pi)^2{R_6\sqrt{g}\over R_1}
\sum_{\vec k\in\Z^4 \ne \vec 0}\big(-{1\over 2} \widetilde G_L^{66}g^{ii'}
k_6 k_6 + {1\over 2} (g^{ii'}g^{jj'}- g^{ij'}g^{ji'}) k_jk_{j'} \big)
(a_{\vec k\,i} a^\dagger_{\vec k\,i'}
+ a^\dagger_{\vec k\,i}a_{\vec k\,i'}),
\end{align} 
with the delta function
\begin{align}
\int {d^4\theta\over (2\pi)^4} e^{i(k_i-k'_i)\theta^i}=\delta_{\vec k,\vec k'},
\end{align}
and where $k_6$ is given in $(\ref{k6})$.
From the on-shell and transverse conditions (\ref{5dg}), \hfill\break
$\widetilde G_L^{66} k_6k_6 +|k|^2 =0,$ and $k^ia_{\vec k\, i} =
k^ia^\dagger_{\vec k\, i}=0,$ so the time-dependence of
$H_c$ on $\theta^6$ cancels and
\begin{align}
H_c &=  (2\pi)^2 {R_6\sqrt{g}\over R_1}\sum_{\vec k\in\Z^4 \ne \vec 0}
g^{ii'} |k|^2 \;
(a_{\vec k\,i} a^\dagger_{\vec k\,i'}
+ a^\dagger_{\vec k\,i}a_{\vec k\,i'}).
\end{align} 
Similarly the momenta from (\ref{Pq}) become
\begin{align}
P_i&= -  {R_6\sqrt{g}\over R_1}
\widetilde G_L^{66} g^{jj'} (2\pi)^2\sum_{\vec k\in\Z^4 \ne \vec 0}
k_6 k_i \,\big(a_{\vec k\,j'}a^\dagger_{\vec k\,j}
+ a^\dagger_{\vec k\,j'}a_{\vec k\,j}\big).
\end{align}
Then
\begin{align}
- H_c + i\gamma^i P_i & = \mp 
\sqrt{-\widetilde G_L^{66}}  {R_6\sqrt{g}\over R_1}\;(2\pi)^2
\sum_{\vec k\in\Z^4 \ne \vec 0}  |k| \; \big ( \pm {|k| \over 
\sqrt{-\widetilde G_L^{66}}}
+ i\gamma^i k_i \big) g^{jj'}
\big (a_{\vec k\,j} a^\dagger_{\vec k\,j'}
+ a^\dagger_{\vec k\,j}a_{\vec k\,j'}\big)\cr 
&= \mp i 
\sqrt{-\widetilde G_L^{66}}  {R_6\sqrt{g}\over R_1}\;(2\pi)^2
\sum_{\vec k\in\Z^4 \ne \vec 0} |k| \; \big( \pm i
{\sqrt{-\widetilde G_L^{66}}\over \widetilde G_L^{66}} \; |k|
+ \gamma^i k_i \big) g^{jj'}
\big (a_{\vec k\,j} a^\dagger_{\vec k\,j'}
+ a^\dagger_{\vec k\,j}a_{\vec k\,j'}\big).
\end{align}
Since we are using a Lorentzian signature metric at this point,
$-\widetilde G_L^{66} >0$. 
Then rewriting in terms of a real Euclidean radius $R_6$,
and making the upper sign choice in (\ref{k6}), we have
\begin{align}
- H_c + i\gamma^i P_i & = - i {1\over R_6} \; {R_6\sqrt{g}\over R_1}
\,(2\pi)^2
\sum_{\vec k\in\Z^4 \ne \vec 0} |k| \big( - i R_6 |k|
+\gamma^i k_i \big)  g^{jj'}
\big (a_{\vec k\,j} a^\dagger_{\vec k\,j'}
+ a^\dagger_{\vec k\,j}a_{\vec k\,j'}\big).
\end{align}
Inserting the polarizations as $a_{\vec k\,i} = f_i^\kappa a^\kappa_{\vec k}
$ and $a^\dagger_{\vec k\,i} = f_i^{\lambda\ast}
a^{\lambda\dagger}_{\vec k}$ from (\ref{polarize}) in the
commutator (\ref{aacom}) gives
\begin{align}
[a_{\vec k\,i}, a^\dagger_{\vec k'\,j}]
&={R_1\over R_6\sqrt{g}} {R_6 \over |k|}
{1\over 2 (2\pi)^2} \Big( g_{ij} - {k_ik_j \over |k|^2}\Big)
\delta_{\vec k,\vec k'}
= f_i^\kappa f_j^{\lambda\ast}
[a^\kappa_{\vec k}, a^{\lambda\dagger}_{\vec k}], 
\end{align}
where we choose the normalization
\begin{align}
[a^\kappa_{\vec k}, a^{\lambda\dagger}_{\vec k'}]
= \delta^{\kappa\lambda} \delta_{\vec k,\vec k'}.
\label{akappa}\end{align}
Then the polarization vectors satisfy
\begin{align}
f_i^\kappa f_j^{\lambda\ast} \delta^{\kappa\lambda}
&={R_1\over \sqrt{g}\,|k|} 
{1\over 2 (2\pi)^2} \Big( g_{ij} - {k_ik_j \over |k|^2}\Big),\qquad
g^{jj'}  f_j^\kappa f_{j'}^{\lambda\ast} \delta^{\kappa\lambda} =
{R_1\over \sqrt{g}\, |k|} 
{1\over 2 (2\pi)^2} \cdot 3,\cr
g^{jj'}  f_j^\kappa f_{j'}^{\lambda\ast} &= \delta^{\kappa\lambda}
{R_1\over \sqrt{g}\, |k|}  {1\over 2 (2\pi)^2}.
\nonumber\end{align}
So the exponent in (\ref{wholepf}) is given by 
\begin{align}
-H_c + i\gamma^i P_i &= - i {1\over R_6} \; {R_6\sqrt{g}\over R_1}(2\pi)^2
\sum_{\vec k\in\Z^4 \ne \vec 0} |k| \big( - i R_6 |k|
+ \gamma^i k_i \big)  g^{jj'}
\big ( 2 a^\dagger_{\vec k\,j}a_{\vec k\,j'} +
[ a_{\vec k\,j},  a^\dagger_{\vec k\,j'}]\, \big)\cr
& = -i \sum_{\vec k\in\Z^4 \ne \vec 0} \big(
\gamma^i k_i - i R_6 |k|\,\big) \, a_{\vec k}^{\kappa\dagger}
a_{\vec k}^\kappa \quad -{i\over 2}  \sum_{\vec k\in\Z^4 \ne \vec 0}
\big( -i R_6 |k|\,\big)\,\delta^{\kappa\kappa}.
\label{HplusP}\end{align}
Then the partition function is
\begin{align}
Z^{5d, Maxwell} &\equiv tr   \;\exp\{2\pi (- H_c + i\gamma^i P_i)\}
= Z^{5d}_{\rm zero\, modes} \; Z^{5d}_{\rm osc},
\end{align} where from (\ref{HplusP}),
\begin{align}
Z^{5d}_{\rm osc} &
= tr \; e^{ -2\pi i \sum_{\vec k\in\Z^4 \ne \vec 0} \big(
\gamma^i k_i - i R_6 |k|\,\big) \, a_{\vec k}^{\kappa\dagger}
a_{\vec k}^\kappa  \;\;  - \, \pi R_6 \sum_{\vec k\in\Z^4 \ne \vec 0}
|k| \,\delta^{\kappa\kappa}}.
\label{ZHplusP}\end{align}

\section{\bf Comparison of Oscillator Traces $Z^{5d}_{\rm osc}$  
and $Z^{6d}_{\rm osc}$}

In order to compare the partition functions of the two theories, 
we first review the calculation for the 6d chiral field from \cite{DN}
setting the angles between the circle and five-torus $\alpha, \beta^i =0$. 
The oscillator trace is evaluated by rewriting (\ref{6dpf}) as
\begin{align} -2\pi R_6{\cal H} +i2\pi\gamma^i\P_i&= 
{i\pi\over 12}\int_0^{2\pi} d^5\theta H_{lrs}
\epsilon^{lrsmn} H_{6mn}
= {i\pi\over 2}\int_0^{2\pi} d^5\theta \sqrt{-G}
H^{6mn} H_{6mn}\cr
&={-i\pi \int_0^{2\pi} d^5\theta
(\Pi^{mn} H_{6mn} + H_{6mn} \Pi^{mn})}
\label{chone}\end{align}
where the definitions
$H^{6mn} = {1\over 6\sqrt{-G}}\epsilon^{mnlrs} H_{lrs}$ and
$H_{6mn}= {1\over 6 \sqrt{-G} G^{66}}\epsilon_{mnlrs}H^{lrs}$ follow from
the self-dual equation of motion (\ref{sd3form}).
$\Pi^{mn}(\vec\theta,\theta^6)$,
the field conjugate to  $B_{mn}(\vec\theta,\theta^6)$ is defined
from the Lagrangian
for a general (non-self-dual) two-form
$I_6=\int d^6\theta (-{\sqrt{-G}\over 24})H_{LMN} H^{LMN}$, so
$\Pi^{mn}\nobreak={\textstyle{\delta I_6\over \delta \partial_6
B_{mn}}}  = -{\sqrt{-G}\over 4} H^{6mn}\,.$
The commutation relations of the two-form
and its conjugate field $\Pi^{mn}(\vec\theta,\theta^6)$ are
\begin{align}
[\Pi^{rs}(\vec\theta,\theta^6),
B_{mn}(\vec\theta',\theta^6)]
=& -i\delta^5 (\vec\theta - \vec\theta')
(\delta^r_m\delta^s_n - \delta^r_n\delta^s_m),\cr
[\Pi^{rs}(\vec\theta,\theta^6),
\Pi^{mn}(\vec\theta',\theta^6)]
=& [B_{rs}(\vec\theta,\theta^6),
B_{mn}(\vec\theta',\theta^6)] =0.\nonumber\end{align}
{}From the Bianchi identity $\partial_{[L}H_{MNP]} = 0$ and the fact
that (\ref{sd3form}) implies $\partial^L H_{LMN} =0$, then
a solution to (2.1) is given by a solution to the homogeneous equations
$\partial^L\partial_L B_{MN}=0,$ $\partial^L B_{LN} =0\,.$
These have a plane wave solution
\begin{align}B_{MN}(\vec\theta,\theta^6)
&=f_{MN}(p) e^{ip\cdot\theta} + (f_{MN}(p) e^{ip\cdot\theta})^\ast;
\qquad G^{LN}p_L p_N =0\,;\qquad p^L f_{LN} =0;\label{ceom}\end{align}
and quantum tensor field expansion
\begin{align}
B_{mn} (\vec\theta, \theta^6) = {\rm zero\, modes} \quad +
\sum_{\vec p= p_l \in {\cal Z}^5\ne \vec 0}
( f_{mn}^\kappa b_{\vec p}^\kappa e^{ip\cdot\theta}
+ f_{mn}^{\kappa\ast} b_{\vec p}^{\kappa\dagger} e^{-ip^\ast\cdot\theta})
\end{align}
for the three physical polarizations of the 6d chiral two-form 
\cite{DN}, $1\le\kappa\le 3$.
Because oscillators with different polarizations commute, 
each polarization can be treated separately and the result then cubed.
Without the zero mode term, 
\begin{align}
B_{mn}(\vec\theta, \theta^6) =
\sum_{\vec p\ne 0} ( b_{\vec p mn}\,e^{ip\cdot\theta}
+ b_{\vec p mn}^{\dagger} e^{-ip^\ast\cdot\theta})\,,\end{align}
for  $b_{\vec pmn}= f_{mn}^1 b_{\vec p}^1$ for example, with a similar
expansion for $\Pi^{mn}(\vec\theta, \theta^6)$ in terms of 
$c^{6mn\dagger}_{\vec p}$.
From (\ref{ceom}) the momentum $p_6$ is
\begin{align}
p_6 =  -\gamma^ip_i -iR_6\sqrt{g^{ij} p_ip_j + {p_1^2\over R_1^2}}.
\end{align}
For the gauge choice $B_{6n}=0$, the exponent (\ref{chone}) becomes 
\begin{align}
&-i\pi (2\pi)^5
\sum_{\vec p=p_l\in {\cal Z}^5\ne 0} i p_6 ({\cal C}_{\vec p}^{6mn\dagger}
B_{\vec p mn} +
B_{\vec p mn} {\cal C}_{\vec p}^{6mn\dagger} )\cr
&= -2 i \pi\sum_{\vec p \ne 0} p_6
{\cal C}_{\vec p}^{\kappa\dagger} B_{\vec p}^\lambda
f^{\kappa mn}(p) f^\lambda_{mn}(p)
- i\pi \sum_{\vec p \ne 0} p_6 f^{\kappa mn}(p) f^\kappa_{mn}(p)\cr
&=  -2 i \pi\sum_{\vec p \ne 0} p_6
{\cal C}_{\vec p}^{\kappa\dagger} B_{\vec p}^\kappa
-i \pi \sum_{\vec p \ne 0} p_6 \delta^{\kappa\kappa},
\end{align}
with $B_{\vec p \,mn} \equiv  b_{\vec p \,mn} + b_{-\vec p \,mn}^\dagger $,
${\cal C}_{\vec p}^{6mn\dagger} \equiv  
c_{-\vec p}^{6mn} + c_{\vec p}^{6mn\dagger}$.
The polarization tensors have been restored where $1\le\kappa,\lambda\le 3$ and
the oscillators $B_{\vec p}^\kappa, {\cal C}_{\vec p}^{\lambda\dagger}$ 
satisfy the commutation relation
\begin{align}
[B_{\vec p}^\kappa, {\cal C}_{\vec p}^{\lambda\dagger}]= 
\delta^{\kappa\lambda} \; \delta_{\vec p,\vec p'}.
\label{BCcr}\end{align}
So restricting the manifold to a circle times a five-torus in \cite{DN} 
we have
\begin{align}
&-2\pi R_6{\cal H} +i2\pi\gamma^iP_i\cr&=
-2i\pi \sum_{\vec p \in {\cal Z}^5\ne 0} 
\Big(-\gamma^ip_i -iR_6\sqrt{g^{ij} p_ip_j + {p_1^2\over R_1^2}}\Big)
{\cal C}_{\vec p}^{\kappa\dagger} B_{\vec p}^\kappa
-\pi R_6 \sum_{\vec p  \in {\cal Z}^5} 
\sqrt{g^{ij} p_i p_j + {p_1^2\over R_1^2}}\;
\delta^{\kappa\kappa}
\label{HPexp}\end{align}
The oscillator trace (\ref{6dpf}) is
\begin{align}
Z^{6d}_{\rm osc} &= tr \, e^{-t{\cal H} + i2\pi\gamma^iP_i}
= tr \, e^{-2i\pi \sum_{\vec p \ne 0} p_6
{\cal C}_{\vec p}^{\kappa\dagger} B_{\vec p}^\kappa
-\pi R_6 \sum_{\vec p } \sqrt{g^{ij}p_ip_j + {p^2_1\over R_1^2}}\;\,
\delta^{\kappa\kappa}},\cr 
Z^{6d, chiral} &= Z^{6d}_{\rm zero\,modes}\cdot\bigl ( e^{-\pi R_6 
\sum_{\vec n \in {\cal Z}^5}
\sqrt{g^{ij} n_i n_j + {n^2_1\over R_1^2}}}\, \prod_{\vec n \in {\cal Z}^5
\ne 0}
{1\over 1-e^{-2\pi R_6\sqrt{g^{ij} n_i n_j + {n^2_1\over R_1^2}}
\; +i2\pi\gamma^in_i}}\bigl)^3.\cr
\label{f6doscpf}\end{align}
\vskip-10pt
Regularizing the vacuum energy as in \cite{DN}, the chiral 
field partition function (\ref{6dpf}) becomes
\begin{align}
Z^{6d, chiral}&
= Z^{6d}_{\rm zero\,modes} \cdot
\Bigl ( e^{ R_6 \pi^{-3} \sum_{\vec n\ne \vec 0} {\sqrt{G_5}\over
(g_{ij} n^in^j + R_1^2 (n^1)^2)^{\,3}}}\,
\prod_{\vec n\in {\cal Z}^5\ne \vec 0}
{\textstyle 1\over{1- e^{-2\pi R_6 \sqrt{g^{ij} n_in_j + {(n_1)^2\over R_1^2}} 
+ i 2\pi \gamma^i n_i}}}\Bigr )^3,
\label{oscb}\end{align}
\vskip-10pt
where $Z^{6d}_{\rm zero\,modes}$ is given in (\ref{chzm}).
Lastly we compute the 5d Maxwell partition function (\ref{wholepf})
from (\ref{ZHplusP}),
\begin{align}
Z^{5d, Maxwell}&=  Z^{5d}_{\rm zero\;modes} \cdot
\, tr \; e^{-2i\pi\sum_{\vec k\ne \vec 0} (\gamma^i k_i
-iR_6\sqrt{g^{ij}k_ik_j})
\, a_{\vec k}^{\kappa\dagger} \,a_{\vec k}^{\kappa}
- \pi\sum_{\vec k\ne \vec 0}
(R_6\sqrt{g^{ij} k_ik_j}) \, \delta^{\kappa\kappa}},\label{wpf2}
\end{align}
where $\vec k = k_i = n_i\in {\cal Z}^4$ on the torus.
From the standard Fock space argument

$$tr\;\omega^{\sum_p p a^\dagger_p a_p}
=\prod_p\sum_{k=0}^\infty \langle k |\omega^{p a^\dagger_p a_p} | k\rangle
=\prod_p {\textstyle 1\over {1 - \omega^p}},$$ 
we perform the trace on the oscillators,
\begin{align}
Z^{5d}_{\rm osc} &=  
\,  \Big( e^{-\pi R_6\sum_{\vec n\in {\cal Z}^4} \sqrt{g^{ij}n_in_j}}
\; \; \prod_{\vec n\in {\cal Z}^4\ne \vec 0} {1\over 1 - e^{-i2\pi
(\gamma^in_i -i R_6\sqrt{g^{ij} n_in_j})}}\Big)^3,\label{opf}\\
Z^{5d, Maxwell} &=  Z^{5d}_{\rm zero\;modes} \cdot
\,  \Big( e^{-\pi R_6\sum_{\vec n\in {\cal Z}^4} \sqrt{g^{ij}n_in_j}}
\; \; \prod_{\vec n\in {\cal Z}^4\ne \vec 0} {1\over 1 - e^{-2\pi
R_6\sqrt{g^{ij} n_in_j} -2\pi i\gamma^in_i}} \Big)^3,
\label{wpf3}\end{align}
where $Z^{5d}_{\rm zero\;modes}$ is given in (\ref{pfi}).
(\ref{wpf3}) and (\ref{f6doscpf}) are each manifestly 
$SL(4,{\cal Z})$ invariant 
due to the underlying $SO(4)$ invariance we have labeled as $i=2,3,4,5$.
We use the $SL(4,{\cal Z})$ invariant regularization of the
vacuum energy reviewed in Appendix C to obtain
\begin{align}
Z^{5d, Maxwell}&=  Z^{5d}_{\rm zero\;modes} \cdot
\,  \Big( e^{{3 \over 8}R_6\pi^{-2}
\sum_{\vec n\ne 0} {\sqrt{g}\over ({g_{ij}n^in^j})^{5\over 2}}}
\; \; \prod_{\vec n\in {\cal Z}^4\ne \vec 0} {1\over 1 - e^{-2\pi
R_6\sqrt{g^{ij} n_in_j} - 2\pi i \gamma^in_i}} \Big)^3,
\label{wpf4}\end{align}
where the sum is on the original lattice $\vec n = n^i\in {\cal Z}^4\ne \vec 0,$
and the product is on the dual lattice $\vec n = n_i\in {\cal Z}^4\ne \vec 0.$
In Appendix D we prove that the product of the zero mode contribution 
and the oscillator
contribution in (\ref{wpf4}) is $SL(5,{\cal Z})$ invariant.
In (\ref{wpf4again}) we give an equivalent expression,
\begin{align}
Z^{5d, Maxwell}&=  Z^{5d}_{\rm zero\;modes} \cdot
\Big( e^{\pi R_6\over 6 R_2} \prod_{n\ne 0} {1\over
1 - e^{-2\pi  {R_6\over R_2}|n| +2\pi i \gamma^2 n}}\Big )^3
\cr& \hskip10pt \cdot
\Big( \prod_{n_\alpha\in \Z^3\ne (0,0,0)} e^{-2\pi R_6 <H>_{p_\perp}}
\; \; \prod_{n_2\in {\cal Z}} {1\over 1 - e^{-2\pi
R_6\sqrt{g^{ij} n_in_j} + 2\pi i\gamma^i n_i}} \Big)^3,\cr
\label{wpfyetagain}\end{align}
with $ <H>_{p_\perp}$ defined in (\ref{BES5}).
In Appendix D we also prove the $SL(5,\Z)$ invariance of 
the 6d chiral partition function 
(\ref{oscb}), using the equivalent form (\ref{a6dpf}),
\begin{align}
Z^{6d, chiral} &= Z^{6d}_{\rm zero\,modes}\cdot\bigl (e^{\pi R_6\over 6 R_2}
\prod_{n\in\Z\ne 0} {1\over
1 - e^{- 2\pi {R_6\over R_2} |n| + 2\pi i \,\gamma^2 \,n\,)}}\bigr )^3\cr
&\hskip20pt \cdot
\bigl(\prod_{n_\perp\in\Z^4\ne (0,0,0,0)}
e^{-2\pi R_6 < H>^{6d}_{p_\perp}} \prod_{n_2\in {\cal Z}}
{1\over 1-e^{-2\pi R_6\sqrt{g^{ij} n_i n_j + {n^2_1\over R_1^2}}
\; +i2\pi\gamma^in_i}}\bigr)^3\label{wpfmore}\end{align}
with $< H>^{6d}_{p_\perp}$ in (\ref{BES6}).
Thus the partition functions of the two theories are both $SL(5,\Z)$ invariant, 
but they are not equal.

The comparison of the 6d chiral theory on 
$S^1\times T^5$ and the abelian gauge theory on $T^5$ shows the
exponent of the oscillator contribution to the partition function
for the 6d theory  (\ref{HPexp}),
\begin{align}
&-2\pi R_6{\cal H} +i2\pi\gamma^i\P_i\cr&=
-2\pi \sum_{\vec p \in {\cal Z}^5\ne 0}
\Big( -i\gamma^ip_i + R_6\sqrt{g^{ij} p_ip_j + {p_1^2\over R_1^2}}\,\Big)
\; {\cal C}_{\vec p}^{\kappa\dagger} B_{\vec p}^\kappa
-\pi R_6 \sum_{\vec p \in {\cal Z}^5} 
\sqrt{g^{ij} p_i p_j + {p_1^2\over R_1^2}}\;\;
\delta^{\kappa\kappa},
\cr\label{HPexpb}\end{align}
and for the gauge theory 
(\ref{HplusP}),
\begin{align}
- 2\pi H^{5d} + 2\pi i \gamma^i P^{5d}_i &
= -2\pi\, \sum_{\vec k\in {\cal Z}^4 \ne 0}
\Big(i\gamma^i k_i +  R_6\sqrt{g^{ij}k_ik_j} \Big) \,
\, a_{\vec k}^{\kappa\dagger} \,a_{\vec k}^{\kappa}
- \pi R_6 \sum_{\vec k\in {\cal Z}^4}
\sqrt{g^{ij} k_ik_j} \, \delta^{\kappa\kappa},
\label{addon5b}
\end{align}
differ only by the sum on the Kaluza-Klein modes $p_1$ of $S^1$ since 
for the chiral case $\vec p\in {\cal Z}^5$, and for the Maxwell case
$\vec k\in {\cal Z}^4$. 
Both theories have three polarizations, $1\le \kappa\le 3$, and
from (\ref{BCcr}), (\ref{akappa})
the oscillators have the same commutation relations,
\begin{align}
[B_{\vec p}^\lambda, {\cal C}_{\vec p}^{\lambda\dagger}]=
\delta^{\kappa\lambda} \; \delta_{\vec p,\vec p'},\qquad
[a_{\vec k}^\kappa, a_{\vec k'}^{\lambda\dagger} ]
&= \delta^{\kappa\lambda} \; \delta_{\vec k,\vec k'}.
\end{align}
If we discard the
Kaluza-Klein modes $p_1^2$ in the usual limit \cite{Witten} as 
the radius of the circle $R_1$ is 
very small with respect to the radii and angles $g_{ij}, R_6,$ 
of the five-torus, then the oscillator products in 
(\ref{wpfmore}) and (\ref{wpfyetagain}) are equivalent. 
This holds as a precise limit
since we can
separate the product on $ n_\perp = (n_1,n_\alpha)\ne 0_\perp $ 
in (\ref{wpfmore}),
into 
$(n_1=0,n_\alpha\ne (0,0,0))$ and $(n_1\ne 0, \hbox{all}\; n_\alpha)$,
to find at fixed $n_2$,
\begin{align}
&\prod_{n_\perp\in{\cal Z}^4\ne (0,0,0,0)}
{\textstyle 1\over{1- e^{-2\pi R_6 \sqrt{g^{ij} n_in_j + {(n_1)^2\over R_1^2}}
+ 2\pi i\gamma^in_i}}}\cr
&\hskip30pt
= \prod_{n_\alpha\in \Z^3\ne (0,0,0)}
{\textstyle 1\over{1- e^{-2\pi R_6 \sqrt{g^{ij} n_in_j } + 2\pi i \gamma^in_i
}}}\cdot \prod_{n_1\ne 0, n_\alpha\in {\cal Z}^3}
{\textstyle 1\over{1- e^{-2\pi R_6 \sqrt{g^{ij} n_in_j + {(n_1)^2\over R_1^2}}
+ 2\pi i\gamma^in_i}}}.\cr
\end{align}
In the limit of small $R_1$ the last product reduces to unity, thus 
for $S^1$ smaller than $T^5$
\begin{align}
&\prod_{n_\perp\in{\cal Z}^4\ne (0,0,0,0)}
{\textstyle 1\over{1- e^{-2\pi R_6 \sqrt{g^{ij} n_in_j + {(n_1)^2\over R_1^2}}
+2\pi i\gamma^in_i}}}
\rightarrow \prod_{n_\alpha\in {\cal Z}^3\ne (0,0,0)}
{\textstyle 1\over{1- e^{-2\pi R_6 \sqrt{g^{ij} n_in_j } + 2\pi i\gamma^in_i
}}}.
\label{65lim}\end{align} 
Inspecting the regularized vacuum energies $<H>_{p_\perp}$ and
$<H>_{p_\perp}^{6d}$ in (\ref{BES5}),(\ref{BES6}),

\begin{align}
<H>_{p_\perp\ne 0}&=-{\pi}^{-1}\; |p_\perp|
\sum_{n=1}^\infty\cos(p_\alpha\kappa^\alpha 2\pi n)
{K_1(2\pi n R_2 |p_\perp|)\over n}, \quad
\hbox{for}\quad
|p_\perp|\equiv \sqrt{\widetilde g^{\alpha\beta} n_\alpha n_\beta},\cr
<H>^{6d}_{p_\perp\ne 0}&=-{\pi}^{-1}\; |p_\perp|
\sum_{n=1}^\infty\cos(p_\alpha\kappa^\alpha 2\pi n)
{K_1(2\pi n R_2 |p_\perp|)\over n}, \quad
\hbox{for}\quad|p_\perp|\equiv \sqrt{{(n_1)^2\over R_1^2}
+ \widetilde g^{\alpha\beta} n_\alpha n_\beta},
\end{align}
we see they have the same form of spherical Bessel functions, 
but the argument differs by Kaluza-Klein modes.  
Again separating the product on $n_\perp = (n_1,n_\alpha)$ in (\ref{wpfmore}),
into \hfill\break
$(n_1=0,n_\alpha\ne (0,0,0))$ and $(n_1\ne 0, \hbox{all}\; n_\alpha)$
we have
\begin{align}
\prod_{n_\perp\in\Z^4\ne (0,0,0,0)}
e^{-2\pi R_6 < H>^{6d}_{p_\perp}} =
\bigl(\prod_{n_\alpha\in\Z^3\ne (0,0,0)}
e^{-2\pi R_6 < H>_{p_\perp}}\bigr)\cdot
\bigl(\prod_{n_1\ne 0, n_\alpha\in\Z^3}
e^{-2\pi R_6 < H>^{6d}_{p_\perp}}\bigr).
\label{relH}\end{align}
In the limit $R_1\rightarrow 0$, the last product is unity because
for $n_1\ne 0$, \begin{align}
&\lim_{R_1\rightarrow 0}  \sqrt{{(n_1)^2\over R_1^2}+ \widetilde 
g^{\alpha\beta} n_\alpha n_\beta} \sim {|n_1|\over R_1},\cr
&\lim_{R_1\rightarrow 0}  |p_\perp| \; K_1(2\pi n R_2 |p_\perp|)
=\lim_{R_1\rightarrow 0} {|n_1|\over R_1}\: K_1
\big (2\pi n R_2 {|n_1|\over R_1}\big)
=0,
\end{align} 
since $\lim_{x\rightarrow\infty} x K_1(x) \sim \sqrt{x} \;e^{-x} \rightarrow 0.$
\cite{AS}. So (\ref{relH}) leads to 
\begin{align}
\lim_{R_1\rightarrow 0} \prod_{n_\perp\in\Z^4\ne (0,0,0,0)}
e^{-2\pi R_6 < H>^{6d}_{p_\perp}} = 
\prod_{n_\alpha\in\Z^3\ne (0,0,0)} e^{-2\pi R_6 < H>_{p_\perp}}.
\end{align} 
Thus in the limit where the radius of the circle $S^1$ is
small with respect to $T^5$, which is the
limit of weak coupling $g_{5YM}^2$, we have proved
\begin{align}
\lim_{R_1\rightarrow 0} & \prod_{n_\perp\in\Z^4\ne (0,0,0,0)}
e^{-2\pi R_6 < H>^{6d}_{p_\perp}} \prod_{n_2\in {\cal Z}}
{1\over 1-e^{-2\pi R_6\sqrt{g^{ij} n_i n_j + {n^2_1\over R_1^2}}
\; +i2\pi\gamma^in_i}}\cr&
= \prod_{n_\alpha\in \Z^3\ne (0,0,0)} e^{-2\pi R_6 <H>_{p_\perp}}
\; \; \prod_{n_2\in {\cal Z}} {1\over 1 - e^{-2\pi
R_6\sqrt{g^{ij} n_in_j} + 2\pi i\gamma^i n_i}}.
\end{align}
So together with
(\ref{zerocomp}), we have shown the partition functions
of the chiral theory on $S^1\times T^5$ and of Maxwell theory on $T^5$,
which we computed in (\ref{wpfmore}) and (\ref{wpfyetagain}),
are equal only in the weak coupling limit,
\begin{align}
\lim_{R_1\rightarrow 0} \; Z^{6d, chiral} = Z^{5d, Maxwell}.
\end{align}

\section{\bf Discussion and Conclusions}

We have addressed a conjecture of the quantum equivalence between
the six-dimensional conformally invariant $N=(2,0)$ theory
compactified on a circle and the five-dimensional maximally 
supersymmetric Yang-Mills theory. In this paper we consider an abelian case
without supersymmetry when the five-dimensional manifold is a twisted torus.
We compute the partition
functions for the chiral tensor field $B_{LN}$ on $S^1\times T^5$,
and for the Maxwell field $A_{\tilde m}$ on $T^5$. We prove the two
partition functions are each $SL(5,\Z)$ invariant, but are
equal only in the limit of weak coupling
$g^2_{5YM}$, a parameter which is proportional to $R_1$, the radius of 
the circle $S^1$.

To carry out the computations we first
restricted an earlier calculation \cite{DN} of the chiral partition function 
on $T^6$ to $S^1\times T^5$. Then we used an operator quantization
to compute the Maxwell partition on $T^5$ as defined in (\ref{wholepf})
which inserts non-zero $\gamma^i$ as the coefficient of $P^{5d}_i$,
but otherwise quantizes the theory in a 5d Lorentzian signature metric
that has zero angles with its time direction, {\it i.e.} $\gamma^i=0,\;
2\le i\le 5,$
\cite{GSW}. We used this metric and form (\ref{wholepf})
to derive both the zero mode and oscillator contributions.
The Maxwell field theory was thus quantized on $T^5$, 
with the Dirac method of constraints resulting in  
the commutation relations in (\ref{aacom}). 

Comparing the partition function of the Maxwell field on a twisted
five-torus $T^5$ with that of a two-form potential with a self-dual three-form
field strength on $S^1\times T^5$,  where the radius of the circle
is $R_1\equiv g^2_{5YM}/ 4\pi^2$, we find the two theories are not equivalent
as quantum theories, but are equal only  
in the limit where $R_1$ is small relative to the metric parameters of 
the five-torus, a limit which effectively removes the Kaluza-Klein modes 
from the 6d partition sum.
How to incorporate these modes rigorously in the 5d theory, 
possibly interpreted as instantons in the non-abelian version of the 
gauge theory  
with appropriate dynamics remains difficult \cite{Seibergone}-\cite{Bern},
suggesting that the 6d finite conformal $N=(2,0)$ theory on a circle
is an ultraviolet completion of the 5d maximally supersymmetric
gauge theory rather than an exact quantum equivalence. 

Furthermore, it would be compelling
to find how expressions for the partition function of the 6d $N=(2,0)$
conformal quantum theory computed on various manifolds using localization
should reduce to the expression in \cite{DN} in an appropriate limit,
providing a check that localization is equivalent to canonical quantization.
 
\section*{Acknowledgments}
We are grateful to Michael Douglas, Peter Goddard, Sergei Gukov and
Edward Witten for discussions.
LD thanks the Institute for Advanced Study at Princeton for its hospitality.
LD and YS are partially supported by the U.S. Department of Energy, Grant No.
DE-FG02-06ER-4141801, Task A.
\vfill\eject
\appendix

\section{\bf Dirac Method of Quantization with Constraints}

The 5d Maxwell theory on a five-torus with metric (\ref{Lfivemetric}) has
the Hamiltonian (\ref{Hp}),
\begin{align}
H_p&= \int d^4\theta \Big(
 -{2\pi^2 R_1\over R_6\sqrt{g} \widetilde G_L^{66}}\, g_{ii'}\Pi^i\,\Pi^{i'} +
{R_6\sqrt{g}\over 16\pi^2 R_1} g^{ii'}g^{jj'} F_{ij} F_{i'j'}
- \partial_i\Pi^i \,A_6   + \lambda_1\Pi^6\Big),
\label{Hpa}\end{align}
with $\lambda_1$ as a Lagrange multiplier.  
To quantize and derive the commutation relations, we start with
the equal-time {\it canonical
Poisson brackets} 
\begin{align}&
\{\Pi^{\tilde m}(\vec\theta,\theta^6), A_{\tilde n}(\vec\theta',\theta^6)\}
=-\{A_{\tilde n}(\vec\theta',\theta^6), \Pi^{\tilde m}(\vec\theta,\theta^6)\} =
-\delta^4(\vec\theta-\vec\theta')\;\delta^{\tilde m}_{\tilde n},\cr
&\{\Pi^{\tilde m}(\vec\theta,\theta^6),\Pi^{\tilde n}(\vec\theta',\theta^6)\}=
\{ A_{\tilde m}(\vec\theta,\theta^6), A_{\tilde n}(\vec\theta',\theta^6)\}=0.
\label{Poisson}
\end{align}
The constraints are required to be time-independent, so for $\phi^1(\theta)
\equiv \Pi^6(\vec\theta,\theta^6)$,
\begin{align}
\partial_6 \phi^1(\vec\theta,\theta^6) = \{\phi^1(\vec\theta,\theta^6),
H_p\}= -\int d^4\theta' \{ \Pi^6(\theta), A_6(\theta')\}
\; \partial_i\Pi^i(\theta') = \partial_i\Pi^i(\theta)
\approx 0. \label{fc}
\end{align}
Thus the secondary constraint is
\begin{align}
\phi^2(\theta) \equiv \partial_i\Pi^i(\vec\theta,\theta^6)\approx 0,
\end{align}
which is time-independent from the contribution
\begin{align}
&\partial_6 \phi^2(\vec\theta,\theta^6) =
\{\phi^2(\vec\theta,\theta^6), H_p\}= {R_6\sqrt{g}\over 16\pi^2 R_1}
g^{ii'}g^{jj'} \int d^4\theta' \{ \partial_k\Pi^k(\theta),
F_{ij}(\theta') F_{i'j'}(\theta')\} = 0.
\label{a5}\end{align}
The two constraints $\phi^1,\phi^2$ are first class constraints since
they have vanishing Poisson bracket,
\begin{align}
\{\Pi^6(\theta),\partial_i\Pi^i(\theta')\} =0.
\end{align}
We introduce the gauge conditions
\begin{align} \phi^3(\theta) &\equiv A_6(\theta) \approx 0,\qquad
\phi^4(\theta) \equiv \partial^iA_i(\theta) = g^{ij}\partial_jA_i\approx 0.
\end{align}
These convert all four constraints to second class,
{\it i.e.}
all now have at least one non-vanishing Poisson bracket with each other,
where the non-vanishing brackets are
\begin{align}
&\{\phi^1(\theta), \phi^3(\theta')\} =
\{\Pi^6(\theta), A_6(\theta')\} = -\delta^4(\theta-\theta')
= -\{A_6(\theta), \Pi^6(\theta')\},\cr
&\{\phi^2(\theta), \phi^4(\theta')\} =
\{\partial_i\Pi^i(\theta), g^{jj'}\partial_{j'} A_j(\theta')\}
=g^{ij} {\partial\over\partial\theta^i} {\partial\over\partial\theta^j}
\delta^4(\theta-\theta') = - \{ 
g^{jj'}\partial_{j'} A_j(\theta),
\partial_i\Pi^i(\theta')\}.
\cr\label{constraintcom}\end{align}
Furthermore, there are no new constraints since
$\partial_6\phi^A
(\vec\theta,\theta^6) = \{ \phi^A(\vec\theta,\theta^6), H\} \approx 0 $,
when all $\phi^A\approx 0, \; 1\le A\le 4,$ and $\lambda_1 = \partial_6A_6.$
We can write (\ref{constraintcom}) as a matrix\hfill\break
$C^{AB}(\theta,\theta')\equiv
\{\phi^A(\theta),\phi^B(\theta')\},$
\begin{align} C^{AB}&=
\left( \begin{matrix}
0&0&-1& 0\cr
0&0&0&g^{ij} {\partial\over\partial\theta^i}
{\partial\over\partial\theta^j} \cr
1&0&0&0\cr
0&-g^{ij} {\partial\over\partial\theta^i}
{\partial\over\partial\theta^j}&0&0
\end{matrix}\right)\;\delta^4(\theta-\theta').
\end{align}
The inverse matrix is
\begin{align}
(C_{AB})^{-1}&=
\left( \begin{matrix}
0&0&1&0\cr
0&0&0& -{1\over g^{kk'}
{\partial\over\partial\theta^k} {\partial\over\partial\theta^{k'}}}\cr
-1&0&0&0\cr
0& {1\over g^{kk'}
{\partial\over\partial\theta^k} {\partial\over\partial\theta^{k'}}}&0&0
\end{matrix}\right)\;\delta^4(\theta-\theta').
\end{align}
The Dirac bracket is defined to vanish with any constraint,
\begin{align}
\{A_{\tilde m}(\theta),\Pi^{\tilde n}(\theta')\}_D
= \{A_{\tilde m}(\theta),\Pi^{\tilde n}(\theta')\}
- \int d^4\rho d^4&\rho' \Big(
\{A_{\tilde m}(\theta),\Pi^6(\rho)\} C^{-1}_{13}
\{A_6(\rho'),\pi^{\tilde n}(\theta')\}\cr
&+\{A_{\tilde m}(\theta), \partial_i\Pi^i(\rho)\} C^{-1}_{24}
\{\partial^jA_j(\rho'),\Pi^{\tilde n}(\theta')\}\cr
&+\{A_{\tilde m}(\theta),  A_6(\rho)\} C^{-1}_{31}
\{\Pi^6(\rho'),\pi^{\tilde n}(\theta')\}\cr
&+\{A_{\tilde m}(\theta), \partial^jA_j (\rho)\} C^{-1}_{42}
\{\partial_i\Pi^i(\rho'),\Pi^{\tilde n}(\theta')\}.
\big)\cr
\end{align}
So
\begin{align}
\{A_i(\theta),\Pi^j(\theta')\}_D
&= \{A_i(\theta),\pi^j(\theta')\}
- \int d^4\rho d^4\rho' \Big(
\{A_i(\theta), \partial_k\Pi^k(\rho)\} \;C^{-1}_{24}\;
\{\partial^{k'}A_{k'}(\rho'),\pi^j (\theta')\}\big)\cr
&= \Big( \delta^j_i - g^{jj'}
\partial_i{1\over g^{kk'}\partial_k\partial_{k'}}
\partial_{j'}\Big) \;\delta^4 (\theta-\theta'),
\end{align}
where here all $\partial_j$ are with respect to $\theta^j$.
So promoting the Dirac Poisson bracket to a quantum commutator, we derive 
the equal time commutation relations
\begin{align}
[\Pi^j(\vec\theta, \theta^6), A_i(\vec\theta',\theta^6)] &=
- i \Big( \delta^j_i - g^{jj'}
(\partial_i{1\over g^{kk'}\partial_k\partial_{k'}}
\partial_{j'})\Big) \;\delta^4 (\theta-\theta'),
\label{DirCom}\end{align}
and similarly, 
\begin{align}
[ A_i(\vec\theta,\theta^6), A_j(\vec\theta',\theta^6)]=0,
\qquad [\Pi^i(\vec\theta,\theta^6),\Pi^j(\vec\theta',\theta^6)]=0.
\label{othcom}\end{align}
Furthermore we can check explicitly that
Dirac brackets with a constraint vanish, for example
\begin{align}
&\{\Pi^j (\theta), \partial^i A_i(\theta')\}_D
= \{\Pi^j (\theta),  g^{ik}\partial_kA_i(\theta')
- g_{ik}\gamma^k  \Pi^i (\theta')\cr
&= \widetilde G^{jk}_L {\partial\over \partial\theta^{k}}
\delta^4(\theta-\theta') -
\widetilde G^{jl}_L {\partial\over \partial\theta^{l}}
\delta^4(\theta-\theta') = 0 = [\Pi^j (\theta), \partial^i A_i(\theta')],
\end{align}
and
\begin{align}[\partial_j\Pi^j (\theta), A_i(\theta')]&=
\partial_j \Big( \delta^j_i - g^{jj'}
(\partial_i{1\over g^{kk'}\partial_k\partial_{k'}}
\partial_{j'})\Big) \;\delta^4 (\theta-\theta') =0.
\end{align}

\section{\bf Equations of Motion}
\label{appendixb}
We check that the Hamiltonian
gives the correct equations of motion
for $A_6=0$ which are derived from $\cL$ given in (\ref{5dLor}):
\begin{align}
\partial^{\tilde m} F_{\tilde m\tilde n} &=
\partial^{\tilde m} \partial_{\tilde m} A_{\tilde n}
-\partial_{\tilde n} \partial^{\tilde m} A_{\tilde m}\cr
& \Rightarrow g^{ij}\partial_i\partial_j A_k 
+ \widetilde G_L^{66}\partial_6\partial_6 A_k
- \partial_k g^{ij}\partial_j A_i 
= 0,\qquad \hbox{for $\widetilde n=k,$}\cr\label{eoma}\\
&\Rightarrow  
g^{ki'}\partial_{i'}\partial_6 A_k = 0,
\qquad\hbox{for $\widetilde n =6.$}
\label{eoma2}\end{align}
Hamilton's equations are
\begin{align}
\partial_6 A_k(\theta) &= \{A_k(\theta), H_p\}
= - {4\pi^2 R_1\over R_6 \sqrt{g} \widetilde G_L^{66}} g_{ki}\Pi^i(\theta) 
+ \partial_k A_6(\theta),\label{H1}\\
\partial_6\Pi^k(\theta) &= \{\Pi^k(\theta), H_p\}
= {R_6\sqrt{g}\over 4\pi^2 R_1}  g^{ii'}g^{kj'} \partial_i F_{i'j'}(\theta).
\label{H2}\end{align}
where regular Poisson brackets are used to compute the time
evolution as in (\ref{a5}). 
(\ref{H1}) is simply the definition of $\Pi^i$ in (\ref{conjmomL}).
(\ref{H2}) is Faraday's law,
\begin{align}
-\partial_6 F^{6k} =\partial_i F^{ik}
\end{align}
which is (\ref{eoma}).
Gauss' law (\ref{eoma2}) follows from the constraint condition
$\phi^4 = \partial^iA_i \approx 0.$
\section{Regularization of the Vacuum Energy for 5d Maxwell Theory}

The Fourier transform of powers of a radial function is
\begin{align}
|\vec p|^{\alpha -  n} &=
{c_\alpha\over (2\pi)^n}\int d^n y \sqrt{G_n} \; 
e^{-i\vec p\cdot\vec y} {1\over
|\vec y|^\alpha},\qquad{\rm where}\qquad
c_\alpha \equiv {\pi^{n\over 2} 2^\alpha \Gamma({\alpha\over 2})\over
\Gamma({n-\alpha \over 2})}.
\label{powrad}\end{align}
This formula holds by analytic continuation, since for general 
$n,\alpha$, where the  area of the unit sphere $S_{n-2}$ is
\begin{align}
\omega_{n-2} = {2\pi^{n-1\over 2}\over \Gamma({n-1\over 2})}
\equiv \int_0^\pi d\theta_1 d\theta_2\ldots d\theta_{n-3}\;
\sin\theta_1 \,\sin^2\theta_2\ldots \sin^{n-3}\theta_{n-3}\int_0^{2\pi}d\phi,
\end{align}
the Fourier integral is 

\begin{align}
\int d^n y \sqrt{G_n} \; e^{-i\vec p\cdot\vec y} {1\over
|\vec y|^\alpha} &=
\int_0^\infty dy \, y^{n-1-\alpha}\int_0^\pi d\theta \sin^{n-2}\theta
\; e^{-i|\vec p| y \cos\theta}\; \omega_{n-2}\cr
&= \int_0^\infty dy \, y^{n-1-\alpha}\;
{(2\pi)^{n\over 2}\over (|\vec p| y)^{n-2\over 2}}\; J_{n-2\over 2}(|\vec p| y)
\cr
&= |\vec p|^{\alpha-n} \; (2\pi)^{n\over 2}\; {2^{{n\over 2}-\alpha}
\Gamma({n-\alpha\over 2})\over\Gamma({\alpha\over 2})},
\qquad\qquad  
\label{genprad}\end{align}
where the last expression is valid for the integral when
$-{n\over 2}<{n\over 2}-\alpha<{1\over 2},$
but can be analytically continued for all $\alpha\ne -n,-n-1,\ldots$

So expressing $|\vec p|$ in terms of its 4d Fourier transform,
\begin{align}
& |\vec p| = -{3\over 4\pi^2}\int d^4 y \sqrt{g} \; e^{-i\vec p\cdot\vec y}
{1\over |\vec y|^5},\cr
&<H> = {1\over 2}\sum_{\vec p\in {\cal Z}^4} |\vec p| \;e^{i\vec p\cdot \vec x}
|_{\vec x =0} = {1\over 2}
\sum_{\vec p\in {\cal Z}^4} \sqrt{g^{ij}p_ip_j},\label{aone}
\end{align}
we have for the sum on the dual lattice, $p_i\in \Z^4$,
\begin{align}&\sum_{\vec p \in {\cal Z}^4}
|\vec p| e^{i{\vec p} \cdot {\vec x}} 
= -{3\over 4\pi^2} \sqrt{g} \int d^4y  {1\over |{\vec y}|^5}
\sum_{\vec p}e^{i{\vec p}\cdot  ({\vec x} - {\vec y})} \cr
&= -{3\over 4\pi^2} \sqrt{g} \int d^4y  {1\over |{\vec y}|^5}
(2\pi)^4\sum_{{\vec n}\ne 0} \delta^4({\vec x}-{\vec y} +2\pi {\vec
n}) = - 12 \pi^2  \sqrt{g} \sum_{{\vec n}\in {\cal Z}^4\ne 0}
{1\over |{\vec x} + 2\pi{\vec n}|^5}  \label{stop}\end{align}
where the regularization consists of removing the ${\vec n}=0$ term from 
the equality,
\begin{align}\sum_{\vec p\in {\cal Z}^4}e^{i{\vec p}\cdot {\vec x}} = (2\pi)^4
\sum_{\vec n\in {\cal Z}^4} \delta^4({\vec x} +2\pi {\vec n})
\label{stim}\end{align}
and the sum on $\vec n$ is on the original lattice
$\vec n = n^i\in {\cal Z}^4$. The regularized vacuum energy is
\begin{align}<H> = -{3\over 16\pi^3}\sqrt{g} \sum_{\vec n \in {\cal Z}^4\ne 0}
{1\over
(g_{ij}n^in^j)^{5\over 2}} = -6\pi^2\sqrt{g}\sum_{{\vec n}\in\Z^4\ne 0} 
{1\over|2\pi {\vec n}|^5}\,. \label{still}\end{align}
For the discussion of $SL(5,\Z)$ invariance in Appendix D,
it is also useful to write the regularized sum (\ref{still}), 
as 
\begin{align}
<H> &= \sum_{p_\perp\in \Z^3} <H>_{p_\perp} 
= <H>_{p_\perp=0} + \sum_{p_\perp\in\Z^3\ne 0}<H>_{p_\perp},
\label{hsplit}\end{align}
where $p_\perp =p_\alpha \in \Z^3,\;\alpha=3,4,5,$ and 
\begin{align}
<H>_{p_\perp=0} = {1\over 2} \sum_{p_2\in\Z} \sqrt{g^{22}p_2p_2}
= {1\over R_2} \sum_{n=1}^\infty n =  {1\over R_2}\zeta(-1)
= -{1\over 12 R_2}
\label{pherp}\end{align}
by zeta function regularization. For general $p_\perp$, 
we express (\ref{still}) as a sum of terms at fixed transverse momentum
\cite{DN},
\begin{align}
<H>_{p_\perp}&= -6\pi^2 \sqrt{g} {1\over (2\pi)^3}\int d^3 z_\perp 
e^{-ip_\perp\cdot z_\perp} \sum_{\vec n\in\Z^4\ne 0} 
{1\over |2\pi\vec n + z_\perp|^5},
\label{3perp}\end{align}
using the equality for the periodic delta function,
\hfill\break
$\sum_{p_\alpha\in\Z^3} e^{ip\cdot z} = (2\pi)^3 \sum_{n^\alpha\in \Z^3}
\delta^3(\vec z + 2\pi\vec n).$
\; Changing variables $z^\alpha\rightarrow y^\alpha + 2\pi n^\alpha$, 
\hfill\break
(\ref{3perp}) becomes
\begin{align}
<H>_{p_\perp} = -6\pi^2\sqrt{g}{1\over (2\pi)^3}
\int d^3y_{\perp}
e^{-ip_{\perp}\cdot
y_{\perp}} \sum_{n\in{\cal Z}\ne 0} {1\over |2\pi n + y_{\perp}|^5}
\label{newstew}\end{align}
where $n$ is the $n^2$ component on the original lattice, 
and the denominator is
$|2\pi n + y_{\perp}|^2\equiv
[(2\pi n)^2 G_{22} + 2 (2\pi n) G_{2\alpha} y_\perp^\alpha +
y_\perp^\alpha y_\perp^\beta G_{\alpha\beta}] =
[(2\pi n)^2 (R_2^2 + g_{\alpha\beta}\kappa^\alpha\kappa^\beta)
- 2 (2\pi n) g_{\alpha\beta} \kappa^\beta  y_\perp^\alpha +
y_\perp^\alpha y_\perp^\beta g_{\alpha\beta}]$.
We can extract the $p_\perp=0$ part of (\ref{newstew}) to verify
(\ref{pherp}),
\begin{align}<H>_{p_\perp=0} &=
-6\pi^2\sqrt{g} {1\over (2\pi)^3} \sum_{n\in {\cal Z}\ne 0}
\int d^3y_{\perp} {1\over  |2\pi n + y_{\perp}|^5}\cr
&=-6\pi^2\sqrt{g} {1\over (2\pi)^3} \sum_{n\in {\cal Z}\ne 0}
{4\pi\over 3} {1\over (2\pi)^2 R_2^2} {1\over n^2}{1\over \sqrt
{\widetilde g}}= -{\zeta (2)\over 2\pi^2 R_2} = - {1\over 12 R_2}, 
\label{stall}\end{align}
by performing the $y$ integrations.
For general $p_\perp\in\Z^3\ne 0$, (\ref{newstew}) integrates to give the
spherical Bessel functions,
\begin{align}
<H>_{p_\perp\ne 0}&=|p_\perp|^2
R_2 \sum_{n=1}^\infty\cos(p_\alpha\kappa^\alpha 2\pi n)
\big [ K_2(2\pi n R_2 |p_\perp|) - K_0(2\pi n R_2 |p_\perp|)\big]\cr
&=-\pi^{-1}\; |p_\perp|
R_2 \sum_{n=1}^\infty\cos(p_\alpha\kappa^\alpha 2\pi n)
{K_1(2\pi n R_2 |p_\perp|)\over n},
\label{BES5}\end{align}
where $|p_\perp|= \sqrt{\widetilde g^{\alpha\beta} n_\alpha n_\beta}$ 
can be viewed as the mass of three scalar bosons \cite{DN}.

For a $d$-dimensional lattice sum, the general formula used in (\ref{aone})
for regulating the divergent sum is \cite{DN},
\begin{align}
|\vec p| &= 2\pi^{-{d\over 2}} {\Gamma({d+1\over 2})\over \Gamma(-{1\over 2})}\;
\int d^d y \sqrt{G_d} \; e^{-i\vec p\cdot\vec y}
{1\over |\vec y|^{d+1}},\cr
<H> &= {1\over 2}\sum_{\vec p\in {\cal Z}^d} |\vec p| \;e^{i\vec p\cdot \vec x}
|_{\vec x =0} = {1\over 2}
\sum_{\vec p\in {\cal Z}^d} \sqrt{g^{\alpha\beta}p_\alpha p_\beta}\cr
&= 2^d \pi^{d\over 2} {\Gamma({d+1\over 2})\over \Gamma(-{1\over 2})}
\sqrt{G_d} \sum_{\vec n\in Z^d\ne\vec 0} {1\over |2\pi\vec n|^{d+1}}.
\label{done}
\end{align}

\section{$SL(5,\Z)$ invariance}

\leftline{\hskip27pt \it Rewriting the 5d metric (2,3,4,5,6)}
From (\ref{sixmetric}) the metric on the five-torus, for $i,j = 2,3,4,5$, is
\begin{align}
G_{ij}&= g_{ij},\qquad G_{i6}=-g_{ij}\gamma^j,\qquad
G_{66}= R_6^2 + g_{ij}\gamma^i\gamma^j,\cr
\widetilde G_5 &\equiv  \det G_{\tilde m\tilde n}  = R^2_6 \,\det g_{ij}
\equiv R^2_6 \, g.
\end{align}
We can rewrite this metric using $\alpha,\beta = 3,4,5$,
\begin{align}
g_{22}& \equiv R_2^2 + \widetilde g_{\alpha\beta}\kappa^\alpha\kappa^\beta,
\qquad g_{\alpha 2} \equiv -\widetilde g_{\alpha\beta} \kappa^\beta,\qquad
g_{\alpha\beta}\equiv\widetilde g_{\alpha\beta},\qquad
(\gamma^2)\kappa^\alpha - \gamma^\alpha \equiv -\widetilde\gamma^\alpha,\\
\cr
G_{22}&=  R_2^2 + \widetilde g_{\alpha\beta}\kappa^\alpha\kappa^\beta,
\qquad G_{26}= - (\gamma^2) R_2^2 + \widetilde g_{\alpha\beta}\kappa^\beta
\widetilde\gamma^\alpha,\qquad G_{2\alpha} = -\widetilde g_{\alpha\beta}
\kappa^\beta,\cr G_{\alpha\beta} &= \widetilde g_{\alpha\beta},
\hskip70pt
G_{\alpha 6} = -\widetilde g_{\alpha\beta}\widetilde\gamma^\beta,
\hskip40pt G_{66} = R_6^2 + (\gamma^2)^2 R_2^2 + \widetilde g_{\alpha\beta}
\,\widetilde\gamma^\alpha \widetilde\gamma^\beta.
\label{our5dmetric}\end{align}
The $4d$ inverse of $g_{ij}$ is
\begin{align}
g^{\alpha\beta} = \widetilde g^{\alpha\beta} + {\kappa^\alpha\kappa^\beta\over
R_2^2},\qquad g^{\alpha 2} = {\kappa^\alpha\over R_2^2},\qquad
g^{22}={1\over R_2^2},
\end{align} where $\widetilde g^{\alpha\beta}$ is the $3d$
inverse of $\widetilde g_{\alpha\beta}$.
$$ g \equiv \det g_{ij} = R_2^2 \; \det \widetilde g_{\alpha\beta}
\equiv  R_2^2 \; \widetilde g.$$
The line element can be written as
\begin{align}
ds^2 &= R_6^2 (d\theta^6)^2 + \sum_{i,j=2,\dots,5} g_{ij}
(d\theta^i-\gamma^i d\theta^6)(d\theta^j-\gamma^j d\theta^6)\cr
&= R_2^2 (d\theta^2 - (\gamma^2) d\theta^6)^2 + R_6^2 (d\theta^6)^2 \cr
&\hskip15pt
+ \sum_{\alpha,\beta = 3,4,5} \widetilde g_{\alpha\beta}(d\theta^\alpha
-\widetilde\gamma^\alpha d\theta^6 - \kappa^\alpha d\theta^2)\,
(d\theta^\beta
-\widetilde\gamma^\beta d\theta^6 - \kappa^\beta d\theta^2).
\label{5dlineelement}\end{align}
We define \begin{align}\widetilde \tau \equiv \gamma^2 + i {R_6\over R_2}.
\label{5dtau}\end{align}
The 5d inverse is
\begin{align}
&\widetilde G_5^{22} = {|\widetilde\tau|^2\over R_6^2} = \widetilde G_5^{66}
|\widetilde\tau|^2,\qquad \widetilde G_5^{66} = {1\over R^2_6},\qquad
\widetilde G_5^{26} = {\gamma^2\over R_6^2}, \qquad
\widetilde G_5^{2\alpha} = {\kappa^\alpha |\widetilde\tau|^2\over  R_6^2}
+ {\gamma^2\widetilde\gamma^\alpha\over R_6^2},\cr
&\widetilde G_5^{\alpha\beta} = \widetilde g^{\alpha\beta} +
{\kappa^\alpha\kappa^\beta\over R_6^2} |\widetilde\tau|^2
+ {\widetilde\gamma^\alpha \widetilde\gamma^\beta\over R_6^2}
+ {\gamma^2(\widetilde\gamma^\alpha\kappa^\beta +
\kappa^\alpha\widetilde\gamma^\beta )\over R_6^2},\qquad
\widetilde G_5^{6\alpha} ={\gamma^\alpha\over R_6^2}
= {\gamma^2\kappa^\alpha + \widetilde \gamma^\alpha
\over R_6^2}.
\label{invalpha}\end{align}

\noindent {\it Generators of $SL(n,{\cal Z})$}

The $SL(n,{\cal Z})$ unimodular groups can be generated by two
matrices \cite{Coxeter}. For $SL(5,{\cal Z})$ these can be taken to be
$U_1, U_2$,
\begin{align}
U_1 &= \left (\begin{matrix}
0&1&0&0&0\cr
0&0&1&0&0\cr
0&0&0&1&0\cr
0&0&0&0&1\cr
1&0&0&0&0
\end{matrix}\right)\,;\qquad
U_2 = \left (\begin{matrix}
1&0&0&0&0\cr
1&1&0&0&0\cr
0&0&1&0&0\cr
0&0&0&1&0\cr
0&0&0&0&1
\end{matrix}\right),
\end{align}
so that every matrix $M$ in
$SL(5,{\cal Z})$ can be written
as a product $U_1^{n_1} U_2^{n_2} U_1^{n_3}\dots$.
Therefore to prove the $SL(5,\Z)$ invariance of (\ref{wpf4}), we will show
it is invariant under $U_1$ and $U_2$.
Matrices $U_1$ and $U_2$ act on the basis vectors of the
five-torus $\vec\alpha_{\tilde m}$ where $\vec\alpha_{\tilde m}
\cdot\vec\alpha_{\tilde n}\equiv
\alpha_{\tilde m}^{\tilde p} \alpha_{\tilde n}^{\tilde q} G_{\tilde p\tilde q}
= G_{\tilde m\tilde n}$,
\begin{align}
\vec\alpha_{2}&= (1,0,0,0,0)\cr
\vec\alpha_{6}&= (0,1,0,0,0)\cr
\vec\alpha_{3}&= (0,0,1,0,0)\cr
\vec\alpha_{4}&= (0,0,0,1,0)\cr
\vec\alpha_{5}&= (0,0,0,0,1).
\end{align}
For our metric (\ref{our5dmetric}), the $U_2$ transformation
\begin{align}\left(\begin{matrix}\vec\alpha'_2\cr
\vec\alpha'_6\cr
\vec\alpha'_3\cr
\vec\alpha'_4\cr
\vec\alpha'_5
\end{matrix}\right)
=U_2\left(\begin{matrix}\vec\alpha_2\cr
\vec\alpha_6\cr
\vec\alpha_3\cr
\vec\alpha_4\cr
\vec\alpha_5\end{matrix}\right) =
\left (\begin{matrix}
1&0&0&0&0\cr
1&1&0&0&0\cr
0&0&1&0&0\cr
0&0&0&1&0\cr
0&0&0&0&1
\end{matrix}\right)
\end{align} results in
$\vec\alpha'_{2}
\cdot\vec\alpha'_{2}\equiv
{\alpha'}^{\tilde p}_2 {\alpha'}^{\tilde q}_2 G_{\tilde p\tilde q}
= G_{22} = G'_{22}, \;\;$$\vec\alpha'_{2}
\cdot\vec\alpha'_{6}\equiv
{\alpha'}^{\tilde p}_{2}{\alpha'}^{\tilde q}_{6} G_{\tilde p\tilde q}
= G_{22} + G_{26} = G'_{26}$, etc.
So $U_2$ corresponds to
\begin{align} R_2\rightarrow R_2 , \; R_6  \rightarrow R_6, \;\gamma^2
\rightarrow \gamma^2 - 1, \;\kappa^\alpha\rightarrow \kappa^\alpha,\;
\widetilde\gamma^\alpha \rightarrow \widetilde\gamma^\alpha
+ \kappa^\alpha, \;\widetilde g_{\alpha\beta}\rightarrow
\widetilde g_{\alpha\beta},\label{U2transf}\end{align}
or equivalently
\begin{align}
R_6\rightarrow R_6, \; \gamma^2\rightarrow \gamma^2 -1,\;
g_{ij}\rightarrow g_{ij},\; \gamma^\alpha\rightarrow\gamma^\alpha,
\label{transfu2}\end{align}
which leaves invariant the line element (\ref{5dlineelement})
if $d\theta^2\nobreak\rightarrow d\theta^2 - d\theta^6, \,
d\theta^6 \rightarrow d\theta^6,\,
d\theta^\alpha \rightarrow d\theta^\alpha.$
$U_2$ is the generalization of the usual $\widetilde\tau\rightarrow
\widetilde\tau - 1$
modular transformation.
The 4d inverse metric $g^{ij} \equiv \{g^{\alpha\beta}, g^{\alpha 2}, g^{22}\}$
does not change under $U_2$.
It is easily checked that $U_2$ is an invariance of the 5d Maxwell
partition function (\ref{wpf3}) as well as the
chiral partition function (\ref{oscb}). It leaves the zero mode
and oscillator contributions invariant separately. 

The other generator, $U_1$ is related to the 
$SL(2,\Z)$ transformation $\widetilde\tau\rightarrow 
-(\widetilde\tau)^{-1}$ that we discuss as follows:
\begin{align}
U_1 = U' M_4
\label{U1}\end{align}
where $M_4$ is an $SL(4,{\cal Z})$ transformation given by
\begin {align}
M_4 &= \left (\begin{matrix}
0&0&-1&0&0\cr
0&1&0&0&0\cr
0&0&0&1&0\cr
0&0&0&0&1\cr
1&0&0&0&0
\end{matrix}\right)
\label{M4}\end{align}
and $U'$ is the matrix corresponding to the transformation on the
metric parameters (\ref{5dmod}),
\begin {align}
U' &= \left (\begin{matrix}
0&1&0&0&0\cr
-1&0&0&0&0\cr
0&0&1&0&0\cr
0&0&0&1&0\cr
0&0&0&0&1
\end{matrix}\right).
\end{align}
Under $U'$, the metric parameters transform as
\begin{align}
R_2&\rightarrow R_2 |\widetilde\tau|,\quad
R_6\rightarrow R_6  |\widetilde\tau|^{-1},\quad
\gamma^2\rightarrow -\gamma^2 |\widetilde\tau|^{-2},\quad
\kappa^\alpha\rightarrow\widetilde\gamma^\alpha,\quad
\widetilde\gamma^\alpha\rightarrow - \kappa^\alpha,\quad
\widetilde g_{\alpha\beta}\rightarrow \widetilde g_{\alpha\beta}.\cr
\widetilde\tau&\rightarrow -{1\over \widetilde \tau}.
\qquad\qquad\hbox{Or equivalently,}\cr
G_{\alpha\beta}&\rightarrow G_{\alpha\beta},\quad
G_{\alpha 2}\rightarrow G_{\alpha 6},\quad
G_{\alpha 6}\rightarrow -G_{\alpha 2},\quad
G_{22}\rightarrow G_{66},\quad
G_{66}\rightarrow G_{22},\quad
G_{26}\rightarrow - G_{26},\cr
\widetilde G_5^{\alpha\beta} &\rightarrow \widetilde G_5^{\alpha\beta},
\quad \widetilde G_5^{\alpha 2}\rightarrow \widetilde G_5^{\alpha 6},
\quad \widetilde G_5^{\alpha 6}\rightarrow -\widetilde G_5^{\alpha 2},
\quad \widetilde G_5^{22}\rightarrow {\widetilde G_5^{22}
\over |\widetilde\tau|^2},
\quad \widetilde G_5^{66}\rightarrow |\widetilde\tau|^2 \widetilde G_5^{66},
\quad \widetilde G_5^{26}\rightarrow
-\widetilde G_5^{26},
\label{5dmod}\end{align}\vskip-16pt
where $3\le \alpha,\beta\le 5,$ and
\begin{align}
\widetilde \tau&\equiv \gamma^2 + i{R_6\over R_2},\qquad |\widetilde\tau|^2 =
(\gamma^2)^2 + {R^2_6\over R_2^2}.
\label{newtau}\end{align}
The transformation (\ref{5dmod}) leaves invariant the line element
(\ref{5dlineelement}) when $d\theta^2\rightarrow d\theta^6,\hfill\break
d\theta^6\rightarrow -d\theta^2,\,$ $
d\theta^1\rightarrow d\theta^1,\,
d\theta^\alpha\rightarrow d\theta^\alpha.$
The generators have the property
$\det U_1 =1, \hfill\break \det U_2 = 1,\;
\det U' =1, \;\det M_4 = 1.$

Under $M_4$, the metric parameters transform as
\begin{align}
&R_6\rightarrow R_6,\qquad \gamma^2\rightarrow -\gamma^3,\quad
\gamma^\alpha\rightarrow \gamma^{\alpha+1},\quad
g_{\alpha\beta}\rightarrow g_{\alpha+1,\beta+1},\quad
g_{\alpha 2}\rightarrow - g_{\alpha+1, 3},\quad
g_{22}\rightarrow g_{33},\cr
&g^{\alpha\beta}\rightarrow g^{\alpha+1,\beta+1},\quad
g^{\alpha 2}\rightarrow - g^{\alpha+1, 3},\quad g^{22}\rightarrow g^{33},
\quad \det g_{ij}= g, \quad g\rightarrow g.
\qquad \hbox{Or equivalently,}\cr\cr
&G_{\alpha\beta}\rightarrow G_{\alpha+1,\beta+1},\quad
G_{\alpha 2}\rightarrow -G_{\alpha+1,3},\quad
G_{\alpha 6}\rightarrow G_{\alpha+1,6},\quad
G_{22}\rightarrow G_{33},\quad
G_{66}\rightarrow G_{66},\quad G_{26}\rightarrow -G_{36},\cr
&\widetilde G_5^{\alpha\beta}\rightarrow \widetilde G_5^{\alpha+1,\beta +1},
\quad \widetilde G_5^{\alpha 2}\rightarrow -\widetilde G_5^{\alpha+1, 3},
\quad\widetilde G_5^{\alpha 6}\rightarrow \widetilde G_5^{\alpha+1,6},\quad
\widetilde G_5^{22}\rightarrow \widetilde G_5^{33},\quad
\widetilde G_5^{26}\rightarrow -\widetilde G_5^{36},\quad
\widetilde G_5^{66}\rightarrow \widetilde G_5^{66},\cr
&\det \widetilde G_5 = R_6 \;g, \qquad \det \widetilde G_5
\rightarrow \det \widetilde G_5,\label{transfm4}\end{align}
where $3\le\alpha,\beta\le 5$, and $\alpha +1 \equiv 2$ for $\alpha =5$.

We can check that ${Z}^{5d}_{\rm zero\, modes}$ is invariant under $M_4$
given in (\ref{M4}) as follows.
Letting the $M_4$ transformation (\ref{transfm4}) act on
(\ref{pfi}), we find that the three subterms in the exponent
\begin{align}
&-2\pi^3{R_6\sqrt{g}\over R_1}
\Big(
g^{\alpha\alpha'} g^{\beta\beta'}
F_{\alpha\beta} F_{\alpha'\beta'}
+ 4 g^{\alpha\alpha'} g^{\beta 2}
F_{\alpha\beta} F_{\alpha' 2}
+  2 g^{\alpha\alpha'} g^{22}
F_{\alpha 2} F_{\alpha' 2}
- 2 g^{\alpha 2} g^{\alpha' 2}
F_{\alpha 2} F_{\alpha' 2}\Big),\cr
&-\pi{R_1R_6\over \sqrt{g}} m^i g_{ij} m^j,\cr
&\hskip13pt i 4\pi^2 \gamma^i m^j  F_{ij}  
\end{align}\vskip-20pt
are separately invariant under  (\ref{transfm4}), if we replace the
the integers $2\pi F_{ij}\in \Z^6, m^i\in \Z^4$ by
\begin{align}
&2\pi F_{\alpha\beta}\rightarrow 2\pi F_{\alpha+1, \beta+1},
\quad 2\pi F_{\alpha 2}\rightarrow - 2\pi F_{\alpha+1, 3},
\quad m^2\rightarrow - m^3,\qquad m^\alpha\rightarrow m^{\alpha + 1},
\end{align}
where $m^i\equiv {2\pi\sqrt{g}\over R_1R_6}g^{ii'}F_{6i'}$
relabels $(n^7,n^8,n^9,n^{10})=(m^2,m^3,m^4,m^5).$
\hfill\break Therefore under $M_4$, for the zero mode contribution,
\begin{align}
\sum_{n_1,\ldots, n_6, n^7,\ldots n^{10}}
e^{-2\pi H^{5d} + i2\pi\gamma^i P_i^{5d}} 
\rightarrow \sum_{n_1,\ldots, n_6, n^7,\ldots n^{10}}
e^{-2\pi H^{5d} + i2\pi\gamma^i P_i^{5d}}.
\label{inv5dzm}\end{align}
So ${Z}^{5d}_{\rm zero\, modes}$ is invariant under $M_4$.
The origin of this is the $SO(4)$ invariance in the coordinate space labeled
by $i=2,3,4,5.$

Next we show under $U'$ that  ${Z}^{5d}_{\rm zero\, modes}$ transforms to 
$|\widetilde\tau|^3 \,  {Z}^{5d}_{\rm zero\, modes}$.
From (\ref{pfi}), 
\begin{align}
Z^{5d}_{\rm zero\;modes}&= \sum_{n_1\ldots n_6}
\,\exp\{-{2\pi^3}{R_6\sqrt{g}\over R_1} g^{ii'}g^{jj'}F_{ij}F_{i'j'}
\} 
\sum_{m^2\ldots m^5} \exp\{-\pi {R_1R_6\over\sqrt{g}} m^i g_{ij} m^j
+ i 4\pi^2 \gamma^i m^j\, F_{ij} \}\cr
&= \sum_{n_1\ldots n_6}
\,\exp\{-{2\pi^3}{R_6\sqrt{g}\over R_1} g^{ii'}g^{jj'}F_{ij}F_{i'j'}
\}
\sum_{m^2\ldots m^5} \exp\{-\pi m \cdot A^{-1} \cdot m + 2\pi i\,m\cdot x\},
\label{pf2}
\end{align}
where $A^{-1}_{ij} = {R_1R_6\over\sqrt{g}} g_{ij}$ and 
$x_j = 2\pi\gamma^i F_{ij}.$ Using a generalization of the Poisson
summation formula
$$ \sum_{m \in {\cal Z}^p} e^{-\pi m\cdot A^{-1}\cdot m} e^{2\pi im\cdot x}\,
= (\det A)^{1\over 2} \sum_{m \in {\cal Z}^p}
e^{-\pi (m+x)\cdot A\cdot (m+x))}$$
we obtain from (\ref{pf2}),
\begin{align}
Z^{5d}_{\rm zero\;modes}&= (\det A)^{1\over 2}\sum_{n_1\ldots n_6\in \Z^6}
\,\exp\{-{2\pi^3}{R_6\sqrt{g}\over R_1} g^{ii'}g^{jj'}F_{ij}F_{i'j'}
\} \cr
&\hskip25pt\cdot
\sum_{m_2\ldots m_5\in \Z^4} \exp\{-\pi {\sqrt{g}\over R_1R_6} g^{jj'} 
(m_j + \gamma^i2\pi F_{ij}) (m_{j'} + \gamma^{i'} 2\pi F_{i'j'})\}, 
\label{pf3}\end{align}
where 
\begin{align} A^{jj'}={\sqrt{g}\over R_1R_6}  g^{jj'},
\qquad \det A = (\det A^{-1})^{-1} = {g\over (R_1R_6)^4}.
\end{align}
To check how this transforms under $U'$ as given in (\ref{5dmod}), it is
convenient to express (\ref{pf3}) in terms of the metric $\widetilde 
G_5^{\tilde l\tilde m}$ found in (\ref{tildeg5}), 
\begin{align}
Z^{5d}_{\rm zero\;modes}&= {\sqrt{g}\over (R_1R_6)^2} 
\sum_{n_1\ldots n_6\in \Z^6}
\,\exp\{-{\pi\over 2}{R_6\sqrt{g}\over R_1} \widetilde G_5^{ii'}
\widetilde G_5^{jj'}(2\pi F_{ij})(2\pi F_{i'j'})
\} \cr
&\cdot
\sum_{m_2\ldots m_5\in \Z^4} \exp\big\{-2\pi {\sqrt{g}R_6\over R_1} 
\widetilde G_5^{6i'} \widetilde G_5^{jj'} m_{j'} (2\pi F_{ij})
-\pi{R_6\sqrt{g}\over R_11} g^{jj'} m_jm_{j'}\big\}.\;
\label{pf4}\end{align}
Curiously we can identify the exponent in (\ref{pf4}) as the 
Euclidean action, if we relabel the integers $m_i$ by $f_{6i}$, 
and the $2\pi F_{ij}$ by $f_{ij}$;
and neglect the integrations.
In this form it will be easy to study its $U'$ transformation, where
(\ref{pf4}) and (\ref{pfi}) can also be written as
\begin{align}Z^{5d}_{\rm zero\;modes}&= 
{\sqrt{g}\over (R_1R_6)^2}
\sum_{f_{\tilde m\tilde n}\in \Z^{10}}
\exp\big\{ -2\pi{\sqrt{\widetilde G_5}\over 4R_1}
\widetilde G_5^{\tilde m \tilde m'} \widetilde G_5^{\tilde n \tilde n'}
f_{\tilde m\tilde n} f_{\tilde m'\tilde n'}\big\}.
\label{pf5}\end{align}
Under $U'$ from (\ref{5dmod}), the coefficient transforms as 
\begin{align}
U':\qquad {\sqrt{g}\over (R_1R_6)^2} \rightarrow {\sqrt{g}\over (R_1R_6)^2}
\;|\widetilde\tau|^3,
\end{align}
since ${\sqrt{g}\over (R_1R_6)^2} = {R_2\sqrt{\widetilde g}\over (R_1R_6)^2}.$
The Euclidean action for the zero mode computation is invariant under $U'$, 
as we show next by first summing $\tilde m = \{2, \alpha, 6\},$ with
$3\le \alpha\le 5.$

\begin{align}
& -2\pi{\sqrt{\widetilde G_5}\over 4R_1}
\widetilde G^{\tilde m \tilde m'} \widetilde G^{\tilde n \tilde n'}
f_{\tilde m\tilde n} f_{\tilde m'\tilde n'}\cr
& = -{\pi R_2R_6\sqrt{\widetilde g}\over 2 R_1}\Big(
\widetilde G_5^{\alpha\alpha'} \widetilde G_5^{\beta\beta'}
f_{\alpha\beta} f_{\alpha'\beta'}
+ 4 \widetilde G_5^{\alpha\alpha'} \widetilde G_5^{\beta 2}
f_{\alpha\beta} f_{\alpha' 2}
+ 4 \widetilde G_5^{\alpha\alpha'} \widetilde G_5^{\beta 6}
f_{\alpha\beta} f_{\alpha' 6}
+ 2 \widetilde G_5^{\alpha\alpha'} \widetilde G_5^{22}
f_{\alpha 2} f_{\alpha' 2}
\cr
&\hskip80pt 
- 2 \widetilde G_5^{\alpha 2} \widetilde G_5^{\alpha' 2}
f_{\alpha 2} f_{\alpha' 2}
+ 4  \widetilde G_5^{\alpha\alpha'} \widetilde G_5^{26}
f_{\alpha 2} f_{\alpha' 6}
-  4 \widetilde G_5^{\alpha 6} \widetilde G_5^{\alpha' 2}
f_{\alpha 2} f_{\alpha' 6}
+ 4 \widetilde G_5^{\alpha 2} \widetilde G_5^{\alpha' 6}
f_{\alpha \alpha'} f_{26}
\cr
&\hskip80pt 
+ 2  \widetilde G_5^{\alpha\alpha'} \widetilde G_5^{66}
f_{\alpha 6 } f_{\alpha' 6}
-  2 \widetilde G_5^{\alpha 6} \widetilde G_5^{\alpha' 6}
f_{\alpha 6} f_{\alpha' 6}
+ 4 \widetilde G_5^{\alpha 2} \widetilde G_5^{26}
f_{\alpha 2} f_{26}
- 4 \widetilde G_5^{\alpha 6} \widetilde G_5^{22}
f_{\alpha 2} f_{26} \cr
&\hskip80pt + 4 \widetilde G_5^{\alpha 2} \widetilde G_5^{66}
f_{\alpha 6} f_{26}
- 4  \widetilde G_5^{\alpha 6} \widetilde G_5^{26}
f_{\alpha 6} f_{26}
- 2  \widetilde G_5^{26} \widetilde G_5^{26}
f_{26} f_{26}
+ 2  \widetilde G_5^{22} \widetilde G_5^{66}
f_{26} f_{26}\Big).
\cr\label{zmalpha}\end{align}
Letting the $U'$ transformation (\ref{5dmod}) act on (\ref{zmalpha}),
we see (\ref{zmalpha}) changes to  
\begin{align}
&\Big(-2\pi{\sqrt{\widetilde G_5}\over 4R_1}
\widetilde G^{\tilde m \tilde m'} \widetilde G^{\tilde n \tilde n'}
f_{\tilde m\tilde n} f_{\tilde m'\tilde n'}\Big)'\cr 
& = -{\pi R_2R_6\sqrt{\widetilde g}\over 2 R_1}\Big(
\widetilde G_5^{\alpha\alpha'} \widetilde G_5^{\beta\beta'}
f_{\alpha\beta} f_{\alpha'\beta'}
+ 4 \widetilde G_5^{\alpha\alpha'} \widetilde G_5^{\beta 6}
f_{\alpha\beta} f_{\alpha' 2}
- 4 \widetilde G_5^{\alpha\alpha'} \widetilde G_5^{\beta 2}
f_{\alpha\beta} f_{\alpha' 6}
+ {2\over |\widetilde\tau|^2} 
\widetilde G_5^{\alpha\alpha'} \widetilde G_5^{22}
f_{\alpha 2} f_{\alpha' 2} 
\cr
&\hskip80pt 
- 2 \widetilde G_5^{\alpha 6} \widetilde G_5^{\alpha' 6}
f_{\alpha 2} f_{\alpha' 2}
- 4  \widetilde G_5^{\alpha\alpha'} \widetilde G_5^{26}
f_{\alpha 2} f_{\alpha' 6}
+  4 \widetilde G_5^{\alpha 2} \widetilde G_5^{\alpha' 6}
f_{\alpha 2} f_{\alpha' 6}
- 4 \widetilde G_5^{\alpha 6} \widetilde G_5^{\alpha' 2}
f_{\alpha \alpha'} f_{26}\cr
&\hskip80pt 
+ 2|\widetilde\tau|^2\widetilde G_5^{\alpha\alpha'} \widetilde G_5^{66}
F_{\alpha 6 } F_{\alpha' 6}
-  2 \widetilde G_5^{\alpha 6} \widetilde G_5^{\alpha' 2}
f_{\alpha 6} f_{\alpha' 6}
- 4 \widetilde G_5^{\alpha 6} \widetilde G_5^{26}
f_{\alpha 2} f_{26}
+ {4\over |\widetilde\tau|^2}
 \widetilde G_5^{\alpha 2} \widetilde G_5^{22}
f_{\alpha 2} f_{26}\cr
&\hskip80pt + 4 |\widetilde\tau|^2
\widetilde G_5^{\alpha 6} \widetilde G_5^{66}
f_{\alpha 6} f_{26}  
- 4  \widetilde G_5^{\alpha 2} \widetilde G_5^{26}
f_{\alpha 6} f_{26}
- 2  \widetilde G_5^{26} \widetilde G_5^{26} F_{26} F_{26}
+ 2  \widetilde G_5^{22} \widetilde G_5^{66} f_{26} f_{26}\Big).
\cr \label{zmtransf}\end{align}
In the partition sum
$\sum_{f_{\tilde m\tilde n}\in \Z^{10}}\; 
e^{-2\pi\Big({\sqrt{\widetilde G_5}\over 4R_1}
\widetilde G^{\tilde m \tilde m'} \widetilde G^{\tilde n \tilde n'}
f_{\tilde m\tilde n} f_{\tilde m'\tilde n'}\Big)'}$,
we can replace the integers as follows:
$f_{\alpha 2} \rightarrow f_{\alpha 6},\;
f_{\alpha 6} \rightarrow - f_{\alpha 2}.
$ Then using (\ref{invalpha}), we have
\begin{align}
\sum_{f_{\tilde m\tilde n}\in \Z^{10}}\; 
e^{-2\pi\Big({\sqrt{\widetilde G_5}\over 4R_1}
\widetilde G^{\tilde m \tilde m'} \widetilde G^{\tilde n \tilde n'}
f_{\tilde m\tilde n} f_{\tilde m'\tilde n'}\Big)'}
= \sum_{f_{\tilde m\tilde n}\in \Z^{10}}\; 
e^{-2\pi\Big({\sqrt{\widetilde G_5}\over 4R_1}
\widetilde G^{\tilde m \tilde m'} \widetilde G^{\tilde n \tilde n'}
f_{\tilde m\tilde n} f_{\tilde m'\tilde n'}\Big)}.
\end{align}
So we have proved that 
under the $U'$ transformation (\ref{5dmod}),
\begin{align}
{Z}^{5d}_{\rm zero\, modes} (R_2|\widetilde \tau|, R_6
|\widetilde \tau|^{-1}, \widetilde g_{\alpha\beta}, -\gamma^2
|\widetilde \tau|^2, \widetilde\gamma^\alpha, -\kappa^\alpha)
= |\widetilde \tau|^3\; {Z}^{5d}_{\rm zero\, modes} (R_2, R_6,
\widetilde g_{\alpha\beta}, \gamma^2,
\kappa^\alpha, \widetilde\gamma^\alpha);
\label{noma}\end{align}
and thus under the $SL(5,\Z)$ generator $U_1$,
${Z}^{5d}_{\rm zero\, modes}$ transforms to
$|\widetilde\tau|^3 \,  {Z}^{5d}_{\rm zero\, modes}$.
(\ref{noma}) also holds for
${Z}^{6d}_{\rm zero\, modes}$, from (\ref{zerocompa}).
This is sometimes referred to as an $SL(2,\Z)$ anomaly of the 
zero mode partition function, because $U'$ includes the 
$\widetilde\tau\rightarrow -{1\over \widetilde \tau}$ transformation.
Finally we will show how this anomaly is canceled by the oscillator
contribution. The 5d and 6d oscillator contributions are not equal,
as given in (\ref{opf}) and (\ref{f6doscpf}). By inspection 
each is invariant under $M_4$, (\ref{transfm4}).  

\vskip20pt
\leftline{\it $U'$ acts on ${Z}^{5d}_{\rm osc}$}
To derive how $U'$ acts on ${Z}^{5d}_{\rm osc}$,
we first separate the product on
$\vec n = (n,n_\alpha)\ne \vec 0$ into a product
on (all $n$, but $n_\alpha\ne\nobreak (0,0,0)$)
and on ($n\ne 0$, $n_\alpha = (0,0,0))$.
Then using the regularized vacuum energy
(\ref{still}) expressed as sum over zero and non-zero transverse momenta
$p_\perp= n_\alpha$
in (\ref{hsplit}),(\ref{pherp}),(\ref{BES5}), we find that 
(\ref{opf})  becomes
\begin{align}
Z^{5d, Maxwell}&=  Z^{5d}_{\rm zero\;modes} \cdot
\Big( e^{\pi R_6\over 6 R_2} \prod_{n\ne 0} {1\over
1 - e^{-2\pi  {R_6\over R_2}|n| -2\pi i \gamma^2 n}}\Big )^3
\cr& \hskip10pt \cdot
\Big( \prod_{n_\alpha\in \Z^3\ne (0,0,0)} e^{-2\pi R_6 <H>_{p_\perp}}
\; \; \prod_{n_2\in {\cal Z}} {1\over 1 - e^{-2\pi
R_6\sqrt{g^{ij} n_in_j} - 2\pi i\gamma^i n_i}} \Big)^3.\cr
\label{wpf4again}\end{align}
As in \cite{DN} we observe the middle expression above can be
written in terms of the Dedekind eta function
$\eta(\widetilde \tau) \equiv e^{\pi i\widetilde\tau\over 12}
\prod_{n\in Z\ne 0} (1-e^{2\pi i\widetilde\tau}),$ with
$\widetilde \tau = \gamma^2 + i{R_6\over R_2},$
\begin{align}
\Big( e^{\pi R_6\over 6 R_2} \prod_{n\ne 0} {1\over
1 - e^{-2\pi  {R_6\over R_2}|n| -2\pi i \gamma^2 n}}\Big )^3
= (\eta(\widetilde \tau)\bar\eta(\bar{\widetilde \tau}))^{-3}.
\label{eta5}\end{align}
This transforms under $U'$ in (\ref{5dmod}) as
\begin{align}
(\eta(-{\widetilde \tau}^{-1})\bar\eta(-\bar{\widetilde \tau}^{-1}))^{-3}
 = |\widetilde\tau|^{-3} \;
 (\eta(\widetilde \tau)\bar\eta(\bar{\widetilde \tau}))^{-3},
\label{etatt}\end{align}
where $\eta(-{\widetilde\tau}^{-1}) = (i\widetilde\tau)^{1\over 2}  
\eta(\widetilde\tau).$
In this way the anomaly of the zero modes in (\ref{noma})
is canceled by the massless part of the oscillator partition function
(\ref{etatt}). Lastly we demonstrate the third expression in (\ref{wpf4again})
is invariant under $U'$,
\begin{align}
\Big( \prod_{n_\alpha\in \Z^3\ne (0,0,0)} e^{-2\pi R_6 <H>_\perp}
\; \; \prod_{n_2\in {\cal Z}} {1\over 1 - e^{-2\pi
R_6\sqrt{g^{ij} n_in_j} - 2\pi i\gamma^i n_i}} \Big)^3
= (PI)^{3\over 2}
\label{5mpf}\end{align}
where $ (PI)^{3\over 2}$ is the modular invariant 2d partition function
of three massive scalar bosons of mass $\sqrt{\widetilde g^{\alpha\beta}
n_\alpha n_\beta},$ coupled to a worldsheet gauge field following \cite{DN}. 
From (\ref{wpf3}),
\begin{align}
Z^{5d}_{\rm osc}&=
( e^{-\pi R_6\sum_{\vec n\in {\cal Z}^4} \sqrt{g^{ij}n_in_j}}
\; \; \prod_{\vec n\in {\cal Z}^4\ne \vec 0} {1\over 1 - e^{-2\pi
R_6\sqrt{g^{ij} n_in_j}}} \Big)^3
\end{align}
we can extract for fixed $n_\alpha \ne 0$,
\begin{align}
(PI)^{1\over 2}  &\equiv
e^{-\pi R_6\sum_{n_2\in {\cal Z}} \sqrt{g^{ij}n_in_j}}
\; \; \prod_{n_2\in {\cal Z}} {1\over 1 - e^{-2\pi
R_6\sqrt{g^{ij} n_in_j} +  2\pi i\gamma^i n_i}}\cr
&= \prod_{s\in {\cal Z}} {e^{-{\beta' E\over 2}}\over 1 -
e^{-\beta' E +  2\pi  i(\gamma^2 s + \gamma^\alpha n_\alpha)}}
\qquad \hbox{where}\; s\equiv n_2,\quad E\equiv \sqrt{g^{ij}n_in_j},\quad
\beta'\equiv 2\pi R_6\cr
&=  \prod_{s\in {\cal Z}} {1\over \sqrt{2}\; \sqrt{\cosh{\beta' E}
- \cos{2\pi(\gamma^2 s + \gamma^\alpha n_\alpha)}}}\qquad
\hbox{for}\; n_\alpha\rightarrow - n_\alpha\cr
&= e^{-{1\over 2}\sum_{s\in{\cal  Z}} \big(\ln {[ \cosh\beta' E -
\cos{2\pi(\gamma^2 s + \gamma^\alpha n_\alpha)}]} + \ln{2}\big)}
\equiv  e^{-{1\over 2}\sum_{s\in{\cal  Z}} \;\nu(E)},
\label{newpi}\end{align}
where
\begin{align}
\sum_{s\in {\cal Z}} \nu(E) &\equiv
\sum_{s\in {\cal Z}} \big(\ln {[ \cosh\beta' E -
\cos{2\pi(\gamma^2 s + \gamma^\alpha n_\alpha)}]} + \ln{2}\big)\cr
&=\sum_{s\in {\cal Z}}\sum_{r\in {\cal Z}}
\ln\, {[{4\pi^2\over {\beta'}^2} ( r + \gamma^2 s + \gamma^\alpha n_\alpha)^2
+ E^2]}\label{eexpr}.\end{align}
(\ref{eexpr}) follows in a similar way to steps (B.3)-(B.3) in \cite{DN},
thus confirming its $U'$ invariance due to the modular invariance of the
massive 2d partition function, which we discuss further in the next section.
We can also show directly that
(\ref{eexpr}) is invariant under $U'$,
since
\begin{align}
&E^2 = g^{ij}n_in_j = g^{22}s^2 + 2 g^{2\alpha}sn_\alpha + g^{\alpha\beta}
n_\alpha n_\beta = {1\over R_2^2} (s + \kappa^\alpha)^2 + \widetilde
g^{\alpha\beta} n_\alpha n_\beta,\cr
& {4\pi^2\over {\beta'}^2} ( r + \gamma^2 s + \gamma^\alpha n_\alpha)^2
= {1\over R_6^2} ( r + \widetilde\gamma^\alpha n_\alpha + \gamma^2(s +
\kappa^\alpha  n_\alpha))^2,
\end{align}
then
\begin{align}
&{4\pi^2\over {\beta'}^2} ( r + \gamma^2 s + \gamma^\alpha n_\alpha)^2
+ E^2\cr & =
{1\over R_6^2} ( s + \kappa^\alpha n_\alpha)^2 \, |\widetilde\tau|^2
+ {1\over R_6^2} ( r + \widetilde \gamma^\alpha n_\alpha)^2
+ {2\gamma^2\over R_6^2} (  r + \widetilde \gamma^\alpha n_\alpha)
( s + \kappa^\alpha n_\alpha) +  \widetilde g^{\alpha\beta} n_\alpha n_\beta.
\cr\label{argln}\end{align}
So we see the transformation $U'$ given in (\ref{5dmod}) leaves
(\ref{argln}) invariant if $s\rightarrow r$ and $r\rightarrow -s$.
Therefore (\ref{eexpr}) is invariant under $U'$, so that
$(PI)^{1\over 2}$ given in  (\ref{newpi}) is invariant under $U'$.
 
In this way,
we have established invariance under $U_1$ and $U_2$, and thus proved 
the partition function for the 5d Maxwell theory
on $T^5$, given alternatively by (\ref{wpf4}) or (\ref{wpf4again}), 
is invariant under $SL(5,\Z)$, the mapping class group of $T^5$.
\vfill\eject

\leftline{\it $U'$ acts on $Z_{osc}^{6d}$}
For the 6d chiral theory on $S^1\times T^5$, the regularized vacuum energy 
from (\ref{oscb}) or (\ref{done}),
\begin{align}
<H>^{6d}&= -32\pi^2 \sqrt{G_5}\sum_{\vec n\ne \vec 0} {1\over (2\pi)^6
(g_{ij} n^in^j + R_1^2 (n^1)^2)^{\,3}}
\end{align}
can be decomposed similarly to (\ref{hsplit}),
\begin{align}
<H>^{6d} &= \sum_{p_\perp\in \Z^3} <H>^{6d}_{p_\perp}
\;= \; <H>^{6d}_{p_\perp=0} + \sum_{p_\perp\in\Z^3\ne 0}<H>^{6d}_{p_\perp},
\label{h6split}\end{align}
where 
\begin{align}
<H>^{6d}_{p_\perp} = -32\pi^2\sqrt{G_5}{1\over (2\pi)^4}
\int d^4y_{\perp}
e^{-ip_{\perp}\cdot
y_{\perp}} \sum_{n^2\in{\cal Z}\ne 0} {1\over |2\pi n^2 + y_{\perp}|^6},
\label{new6stew}\end{align}
with denominator $ |2\pi n^2 + y_{\perp}|^2 =
G_{22} (2\pi n^2)^2 + 2(2\pi n^2)G_{2k}y_\perp^k+G_{kk'}y_\perp^k y_\perp^{k'},$
\begin{align}
<H>^{6d}_{p_\perp=0}&= -{1\over 12 R_2},\cr
<H>^{6d}_{p_\perp\ne 0}& = 
|p_\perp |^2 R_2 \sum_{n=1}^\infty
{\rm cos}(p_\alpha\kappa^\alpha 2\pi n)
\big [K_2(2\pi n R_2 |p_\perp |) - K_0(2\pi n R_2 |p_\perp |)\big]\cr
&=-\pi^{-1}\; |p_\perp|
R_2 \sum_{n=1}^\infty\cos(p_\alpha\kappa^\alpha 2\pi n)
{K_1(2\pi n R_2 |p_\perp|)\over n},
\label{BES6}\end{align}
where $p_\perp = (p_1,p_\alpha) = n_\perp = (n_1,n_\alpha) 
= (n_1,n_3,n_4,n_5) \in {\cal Z}^4$, $|p_\perp| = \sqrt{{(n_1)^2\over R_1^2}
+ \widetilde g^{\alpha\beta} n_\alpha n_\beta}.$

The $U'$ invariance of (\ref{oscb}) follows when 
 we separate the product on $\vec n\in {\cal Z}^5\ne
\vec 0$ into a product on ($n_2\ne 0,\; n_\perp \equiv (n_1,n_3,n_4,n_5)
= (0,0,0,0),$) and on
(all $n_2$, but $n_\perp = (n_1,n_3,n_4,n_5)\ne (0,0,0,0)$). Then
\begin{align}
Z^{6d, chiral} &= Z^{6d}_{\rm zero\,modes}\cdot\bigl (e^{\pi R_6\over 6 R_2}
\prod_{n_2\in\Z\ne 0} {1\over
1 - e^{2\pi i \,(\gamma^2 \,n_2 + {i{R_6\over R_2} |n_2|\,)}}}\bigr )^3\cr
&\hskip20pt \cdot
\bigl(\prod_{n_\perp\in\Z^4\ne (0,0,0,0)}
e^{-2\pi R_6 < H>^{6d}_{p_\perp}} \prod_{n_2\in {\cal Z}}
{1\over 1-e^{-2\pi R_6\sqrt{g^{ij} n_i n_j + {n^2_1\over R_1^2}}
\; +i2\pi\gamma^in_i}}\bigr)^3\cr
&= Z^{6d}_{\rm zero\,modes}\cdot\bigl ( \eta(\widetilde \tau)
\;\bar\eta (\bar{\widetilde \tau})\bigr)^{-3}\cr
&\hskip20pt
\cdot\bigl(\prod_{(n_1,n_3,n_4,n_5)\in\Z^4\ne (0,0,0,0)}
e^{-2\pi R_6 < H>^{6d}_{p_\perp}}
\prod_{n_2\in {\cal Z}}{1\over 1-e^{-2\pi R_6\sqrt{g^{ij} n_i n_j
+ {n^2_1\over R_1^2}} \; +i2\pi\gamma^in_i}}\bigr)^3,\cr
\label{a6dpf}\end{align}
where $\widetilde \tau = \gamma^2 + i{R_6\over R_2}$.
So from the previous section together with 
(\ref{zerocomp}), $U'$ leaves invariant
\begin{align} Z^{6d}_{\rm zero\,modes}\cdot\bigl ( \eta(\widetilde \tau)
\;\bar\eta (\bar{\widetilde \tau})\bigr)^{-3}. 
\end{align} 
The part of the 6d partition function (\ref{a6dpf}) at fixed $n_\perp\ne 0$,
\begin{align}
e^{-2\pi R_6 < H>_{n_\perp\ne 0}}
\prod_{n_2\in {\cal Z}}{1\over 1-e^{-2\pi R_6\sqrt{g^{ij} n_i n_j
+ {n^2_1\over R_1^2}} \; +i2\pi\gamma^in_i}}
\label{MB}\end{align}
corresponds to massive bosons on a two-torus and is
invariant under the $SL(2,{\cal Z})$ transformation $U'$
given in (\ref{5dmod}), as follows \cite{DN}.
Each term with fixed $n_{\perp}\ne 0$ given in
(\ref{MB})
is the square root of
the partition function on $T^2$ (in the directions 2,6)
of a massive complex scalar
with $m^2 \equiv G^{11} n_1^2 + \widetilde g^{\alpha\beta}n_\alpha n_\beta$,
$3\le \alpha,\beta \le 5$,
that couples to a constant gauge field
$A^\mu \equiv i G^{\mu i}  n_i$ with $\mu,\nu ={2,6};\,i,j = 1,3,4,5$.
The metric on $T^2$ is $h_{22} = R_2^2\, , h_{66} = R_6^2 +
(\gamma^2)^2 R_2^2\, , \, h_{26} = -\gamma^2 R_2^2$. Its inverse is
$h^{22}= {1\over R_2^2} + {(\gamma^2)^2\over R_6^2}$, $h^{66}={1\over R_6^2}$
and $h^{26}={\gamma^2\over R_6^2}$. 
The manifestly $SL(2,{\cal Z})$ invariant {\it path integral}
on the two-torus is
\begin{align} {\rm P.I.}
&=\int d\phi \,d\bar\phi \,\,
e^{-\int_0^{2\pi} d\theta^2 \int_0^{2\pi} d\theta^6\,\,
h^{\mu\nu}(\partial_\mu + A_\mu )\bar\phi
(\partial_\nu - A_\nu) \phi + m^2 \bar\phi \phi}\cr
&=\int d\bar\phi\, d\phi
e^{-\int_0^{2\pi} d\theta^2 \int_0^{2\pi} d\theta^6
\bar\phi (-({1\over R_2^2} + {(\gamma^2)^2\over R_6^2})\partial_2^2
-({1\over R_6})^2\partial_6^2 -2 {\gamma^2\over {R_6^2}}
\partial_2\partial_6 + 2A^2\partial_2 + 2A^6\partial_6
+ G^{11}n_1n_1 + G^{\alpha\beta} n_\alpha n_\beta )\phi}\cr
&=\det\Bigl(
[-( {\scriptstyle{1\over R_2^2}}+({\scriptstyle {\gamma^2\over R_6}})^2 )
\partial_2^2 -({\scriptstyle {1\over R_6}})^2\partial_6^2
\hskip-1pt-\hskip-1pt 
2\gamma^2({\scriptstyle {1\over R_6}})^2\partial_2\partial_6
\hskip-1pt +\hskip-1pt G^{11}n_1n_1+G^{\alpha\beta} n_\alpha n_\beta\hskip-1pt
+\hskip-1pt 2iG^{2\alpha}n_\alpha \partial_2\hskip-1pt
+\hskip-1pt 2iG^{6\alpha}n_\alpha \partial_6 ]\Bigr)^{-1}\cr
&= e^{- \tr \ln \Bigl [
- ( {\scriptstyle{1\over R_2^2}} + ({\scriptstyle {\gamma^2\over R_6}})^2 )
\partial_2^2 - ({\scriptstyle {1\over R_6}})^2\partial_6^2
- 2\gamma^2 ({\scriptstyle {1\over R_6}})^2\partial_2\partial_6
+  G^{11}n_1n_1 + G^{\alpha\beta} n_\alpha n_\beta + 2 iG^{2\alpha}n_\alpha
\partial_2 + 2iG^{6\alpha}n_\alpha \partial_6 \, \Bigr ] }
\cr&=e^{-\sum_{s\in{\cal Z}}\sum_{r\in {\cal Z}} \Bigl [ \ln
({4\pi^2 \over {\beta'}^2}r^2 +
( {\scriptstyle{1\over R_2^2}} + ({\scriptstyle {\gamma^2\over R_6}})^2 ) s^2
+ 2\gamma^2 ({\scriptstyle {1\over R_6}})^2 r s
+ G^{11}n_1n_1 + G^{\alpha\beta} n_\alpha n_\beta
+ 2 G^{1\alpha}n_\alpha \,s + 2G^{6\alpha}n_\alpha\, r )
\, \Bigr ]}\cr
&= e^{-\sum_{s\in{\cal Z}}\nu(E)}
\label{btwo}\end{align}
where from (\ref{sixmetricinverse}), $ G^{11}= {1\over R_1^2},
G^{\alpha\beta} =
g^{\alpha\beta} + {\gamma^\alpha\gamma^\beta\over R_6^2},
G^{2\alpha} = g^{2\alpha} + {\gamma^2\gamma^\alpha\over R_6^2},
G^{6\alpha} = {\gamma^\alpha\over R_6^2},$ and 
$\beta' \equiv 2\pi R_6$, and $\partial_2\phi = -is\phi;\;
\partial_6\phi = -ir\phi$, and $n_2 \equiv s$. The sum on $r$ is
\begin{align}\nu (E) &= 
 \sum_{r\in {\cal Z}} \ln \Bigl [{4\pi^2 \over {\beta'}^2}
(r + \gamma^2 s + \gamma^\alpha n_\alpha)^2
+ E^2 \Bigr ]\,,\label{bthree}\end{align}
with
$E^2 \equiv G^{lm}_5 n_l n_m = G_5^{11} n_1n_1 + G_5^{\alpha\beta}
n_\alpha n_\beta + 2 G^{\alpha 2}_5 n_\alpha n_2 + G_5^{22} n_2 n_2,$ 
and $G_5^{11} = {1\over R_1^2}, \hfill\break G_5^{12}=0, G_5^{1\alpha}=0,
G_5^{2\alpha} =  g^{2\alpha} = {\kappa^\alpha\over R_2^2},
G_5^{22} = g^{22} = {1\over R_2^2},
G_5^{\alpha\beta} = g^{\alpha\beta} = \widetilde g^{\alpha\beta}
+ {\kappa^\alpha\kappa^\beta\over R_2^2}.$
We evaluate the divergent sum $\nu(E)$ on $r$ by

\begin{align}{\partial \nu (E)\over \partial E} & =
\sum_r {2E\over {4\pi^2\over {\beta'}^2 } (r + \gamma^2 s +
\gamma^\alpha n_\alpha )^2
+ E^2} \cr
& = \partial_E\ln \Bigl [ \cosh{\beta' E} -
\cos{2\pi \bigl ( \gamma^2 s + \gamma^\alpha n_\alpha \bigr)} \Bigr ],
\label{bfour}\end{align}
using the sum $\sum_{n\in {\cal Z}}
{2y\over {(2\pi n + z)^2 +y^2}} = {\sinh{y}\over {\cosh y -\cos z}}$.
Then integrating (\ref{bfour}), we choose the integration constant
to maintain modular invariance of (\ref{btwo}), 
\begin{align}\nu (E) = \ln \bigl [ \cosh{\beta' E} -
\cos{2\pi \bigl (\gamma^2 s  + \gamma^\alpha n_\alpha\bigr )} \bigr ]
+ \ln 2 \label{bfive}.\end{align}
It follows for $n_2\equiv s$ we have that (\ref{btwo}) is
\begin{align}({\rm P.I.})^{1\over 2} &=
\prod_{s\in {\cal Z}} {1\over
\sqrt 2\sqrt {\cosh{\beta E} -
\cos{2\pi ( \gamma^2 s + \gamma^\alpha n_\alpha )}}}\cr
&= \prod_{s\in {\cal Z}} {e^{-{\beta E\over 2}}\over
{1 - e^{-\beta E + 2\pi i ( \gamma^2 s + \gamma^\alpha n_\alpha )}}}\cr
&= e^{-\pi R_6 {\sum_{s\in {\cal Z}} \sqrt{G_5^{lm} n_l n_m}}}
\prod_{s\in {\cal Z}}
{1\over
{1 - e^{-2\pi R_6 \sqrt{G_5^{lm} n_l n_m}
+ 2\pi i  \gamma^2 s
+ 2\pi i \gamma^\alpha n_\alpha}}}\,\cr
&= e^{-2\pi R_6 <H>_{n\perp}}
\prod_{n_2\in {\cal Z}}
{1\over
{1 - e^{-2\pi R_6 \sqrt{G_5^{lm} n_l n_m}
+ 2\pi i\gamma^2 n_2
+ 2\pi i \gamma^\alpha n_\alpha}}}\,,
\label{bfinal}\end{align}
which is (\ref{MB}).
Its invariance under $U'$ follows since (\ref{5dmod})
is an  $SL(2,{\cal Z})$ transformation on $T^2$ combined with a
gauge transformation on the 2d gauge field,  
$A_\mu\equiv h_{\mu\nu} i n_i G^{\nu i}$ where $\mu,\nu
= {2,6}$, 
$A_\mu\rightarrow A_\mu + \partial_\mu\lambda $,
and $ \phi \rightarrow e^{i\lambda}$, $\bar\phi\rightarrow e^{-i\lambda}$,
\begin{align}\lambda (\theta^1, \theta^6)
&= \theta^2 \, i (\widetilde \gamma^\alpha - \kappa^\alpha)
- \theta^6 \, i (\widetilde \gamma^\alpha + \kappa^\alpha) \label{LAM}
\end{align}
since $A_2 = i\kappa^\alpha n_\alpha,\;
A_6 = i \,\widetilde\gamma^\alpha \,n_\alpha.$
Hence (\ref{bfinal}) and thus (\ref{MB}) are invariant under $U'$.
So we have proved the 6d partition function for the chiral field on 
$S^1\times T^5$, given by  (\ref{oscb}) or equivalently (\ref{a6dpf}), 
is invariant under $U_1$ and $U_2$ and is hence $SL(5,\Z)$ invariant.

\vfill\eject
\singlespacing

\providecommand{\bysame}{\leavevmode\hbox to3em{\hrulefill}\thinspace}
\providecommand{\MR}{\relax\ifhmode\unskip\space\fi MR }
\providecommand{\MRhref}[2]
{
}
\providecommand{\href}[2]{#2}


\begin{thebibliography}{99}



\bibitem{Douglas}
M.~R.~Douglas,
{\it On D=5 super Yang-Mills theory and (2,0) theory,}
JHEP {\bf 1102}, 011 (2011)
[\href{http://xxx.lanl.gov/abs/1012.2880}
{{\tt arXiv:1012.2880 [hep-th]}}].

\bibitem{Lambert}
N.~Lambert, C.~Papageorgakis and M.~Schmidt-Sommerfeld,
{\it M5-Branes, D4-Branes and quantum 5D super-Yang-Mills,}
JHEP {\bf 1101}, 083 (2011)
[\href{http://xxx.lanl.gov/abs/1012.2882}
{{\tt arXiv:1012.2882 [hep-th]}}];
{\it M-Theory and maximally supersymmetric gauge theories,}
[\href{http://xxx.lanl.gov/abs/1203.4244}
{{\tt arXiv:1203.4244 [hep-th]}}];
N.~Lambert, H.~Nastase and C.~Papageorgakis,
{\it 5D Yang-Mills instantons from ABJM Monopoles,}
Phys.\ Rev.\ D {\bf 85}, 066002 (2012)
[\href{http://xxx.lanl.gov/abs/1111.5619}
{{\tt arXiv:1111.5619 [hep-th]}}].

\bibitem{DN}
L.~Dolan and C.~R.~Nappi,
{\it A modular invariant partition function for the five-brane,}
Nucl.\ Phys.\ B {\bf 530}, 683 (1998)
[\href{http://xxx.lanl.gov/abs/hep-th/9806016}
{{\tt arXiv:hep-th/9806016}}];
{\it The Ramond-Ramond selfdual five form's partition function on T**10,}
Mod.\ Phys.\ Lett.\ A {\bf 15}, 1261 (2000)
[\href{http://xxx.lanl.gov/abs/hep-th/0005074}
{{\tt arXiv:hep-th/0005074}}].

\bibitem{WittenBeasley}
C.~Beasley and E.~Witten,
{\it Non-abelian localization for Chern-Simons theory,}
J.\ Diff.\ Geom.\  {\bf 70}, 183 (2005)
[\href{http://xxx.lanl.gov/abs/hep-th/0503126}
{{\tt arXiv:hep-th/0503126]}}].

\bibitem{Pestun}
V.~Pestun,
{\it Localization of gauge theory on a four-sphere and supersymmetric Wilson
loops,} Commun.\ Math.\ Phys.\  {\bf 313}, 71 (2012)
[\href{http://xxx.lanl.gov/abs/0712.2824}
{{\tt arXiv:0712.2824 [hep-th]}}].

\bibitem{Kapustin}
A.~Kapustin, B.~Willett and I.~Yaakov,
{\it Tests of Seiberg-like duality in three dimensions,}
[\href{http://xxx.lanl.gov/abs/1012.4021}
{{\tt arXiv:1012.4021 [hep-th]}}].

\bibitem{DimofteGaiottoGukov}
T.~Dimofte, D.~Gaiotto and S.~Gukov,
{\it 3-Manifolds and 3d Indices,}
[\href{http://xxx.lanl.gov/abs/1112.5179}
{{\tt arXiv:1112.5179 [hep-th]}}].

\bibitem{GRRY} 
A.~Gadde, L.~Rastelli, S.~S.~Razamat and W.~Yan,
{\it The 4d superconformal index from q-deformed 2d Yang-Mills,}
Phys.\ Rev.\ Lett.\  {\bf 106}, 241602 (2011)
[\href{http://xxx.lanl.gov/abs/1104.3850}
{{\tt arXiv:1104.3850 [hep-th]}}].

\bibitem{Tachikawa}
Y.~Tachikawa,
{\it 4d partition function on $S^1\times S^3$ and 2d Yang-Mills with 
nonzero area,}
[\href{http://xxx.lanl.gov/abs/1207.3497}
{{\tt arXiv:1207.3497 [hep-th]}}].

\bibitem{BG}
D.~Bak and A.~Gustavsson,
{\it M5/D4 brane partition function on a circle bundle,}
JHEP {\bf 1212}, 099 (2012)
[\href{http://xxx.lanl.gov/abs/1209.4391}
{{\tt arXiv:1209.4391}}].

\bibitem{Gustavsson}
A.~Gustavsson,
{\it A preliminary test of abelian D4-M5 duality,}
Phys.\ Lett.\ B {\bf 706}, 225 (2011)
[\href{http://xxx.lanl.gov/abs/1111.6339}
{{\tt arXiv:1111.6339 [hep-th]}}];
{\it On the holomorphically factorized partition function for
abelian gauge theory in six-dimensions,}
Int.\ J.\ Mod.\ Phys.\ A {\bf 17}, 383 (2002)
[\href{http://xxx.lanl.gov/abs/hep-th/0008161}
{{\tt arXiv:hep-th/0008161]}}].

\bibitem{Dijkgraaf}
 R.~Dijkgraaf, E.~P.~Verlinde and M.~Vonk,
{\it On the partition sum of the NS five-brane,}
[\href{http://xxx.lanl.gov/abs/hep-th/0205281}
{{\tt arXiv:hep-th/0205281]}}].

\bibitem{GSW}
M.B. Green, J. H. Schwarz and E. Witten,
$\underline{\rm Superstring\, Theory}$,
vol. I and II, Cambridge University Press: Cambridge, U.K. 1987.
See vol. II p40.

\bibitem{KimyeongLee}
H.~-C.~Kim, S.~Kim, E.~Koh, K.~Lee and S.~Lee,
{\it On instantons as Kaluza-Klein modes of M5-branes,}
JHEP {\bf 1112}, 031 (2011)
[\href{http://xxx.lanl.gov/abs/1110.2175}
{{\tt arXiv:1110.2175 [hep-th]}}];
S.~Bolognesi and K.~Lee,
{\it Instanton partons in 5-dim SU(N) gauge theory,}
Phys.\ Rev.\ D {\bf 84}, 106001 (2011)
[\href{http://xxx.lanl.gov/abs/1106.3664}
{{\tt arXiv:1106.3664 [hep-th]}}].

\bibitem{CollieTong}
B.~Collie and D.~Tong,
{\it The partonic nature of instantons,}
JHEP {\bf 0908}, 006 (2009)
[\href{http://xxx.lanl.gov/abs/0905.2267}
{{\tt arXiv:0905.2267 [hep-th]}}].

\bibitem{Singh}
H.~Singh,
{\it Super-Yang-Mills and M5-branes,}
JHEP {\bf 1108}, 136 (2011)
[\href{http://xxx.lanl.gov/abs/1107.3408}
{{\tt arXiv:1107.3408 [hep-th]}}].

\bibitem{Ho}
P.~-M.~Ho, K.~-W.~Huang and Y.~Matsuo,
{\it A non-abelian self-dual gauge theory in 5+1 dimensions,}
JHEP {\bf 1107}, 021 (2011)
[\href{http://xxx.lanl.gov/abs/1104.4040}
{{\tt arXiv:1104.4040 [hep-th]}}].

\bibitem{Samtleben} 
H.~Samtleben,
{\it Actions for non-abelian twisted self-duality,}
Nucl.\ Phys.\ B {\bf 851}, 298 (2011)
[\href{http://xxx.lanl.gov/abs/1105.3216}
{{\tt arXiv:1105.3216 [hep-th]}}].

\bibitem{ChuKo} 
C.~-S.~Chu and S.~-L.~Ko,
{\it Non-abelian action for multiple five-branes with self-dual tensors,}
JHEP {\bf 1205}, 028 (2012)
[\href{http://xxx.lanl.gov/abs/1203.4224}
{{\tt arXiv:1203.4224 [hep-th]}}].

\bibitem{Lee}
K.~-M.~Lee and J.~-H.~Park,
{\it 5-D actions for 6-D selfdual tensor field theory,}
Phys.\ Rev.\ D {\bf 64}, 105006 (2001)
[\href{http://xxx.lanl.gov/abs/hep-th/0008103}
{{\tt arXiv:hep-th/0008103]}}].

\bibitem{Nekrasov}
N.~Nekrasov,
{\it Instanton partition functions and M-theory},
Japanese Journal of Mathematics, {\bf 4}, 63 (2009)
[\href{http://www.springerlink.com/content/b5052p77nk533m47/}
{{\tt springerlink}}]; N.~A.~Nekrasov,
{\it Seiberg-Witten prepotential from instanton counting,}
Adv.\ Theor.\ Math.\ Phys.\  {\bf 7}, 831 (2004)
[\href{http://xxx.lanl.gov/abs/hep-th/0206161}
{{\tt arXiv:hep-th/0206161}}].

\bibitem{Kallenone}
J.~K\"all\'en, J.~Qiu and M.~Zabzine,
{\it The perturbative partition function of supersymmetric 5D Yang-Mills
theory with matter on the five-sphere,}
[\href{http://xxx.lanl.gov/abs/1206.6008}
{{\tt arXiv:1206.6008 [hep-th]}}].

\bibitem{Kallentwo}
J.~K\"all\'en, J.~A.~Minahan, A.~Nedelin and M.~Zabzine,
{\it $N^3$-behavior from 5D Yang-Mills theory,}
[\href{http://xxx.lanl.gov/abs/1207.3763}
{{\tt arXiv:1207.3763 [hep-th]}}].

\bibitem{Kim} 
H.~-C.~Kim and S.~Kim,
{\it M5-branes from gauge theories on the 5-sphere,}
[\href{http://xxx.lanl.gov/abs/1206.6339}
{{\tt arXiv:1206.6339 [hep-th]}}].

\bibitem{Linander}
H.~Linander and F.~Ohlsson,
{\it (2,0) theory on circle fibrations,}
JHEP {\bf 1201}, 159 (2012)
[\href{http://xxx.lanl.gov/abs/1111.6045}
{{\tt arXiv:1111.6045 [hep-th]}}].

\bibitem{Vandoren}
B.~Haghighat and S.~Vandoren,
{\it Five-dimensional gauge theory and compactification on a torus,}
JHEP {\bf 1109}, 060 (2011)
[\href{http://xxx.lanl.gov/abs/1107.2847}
{{\tt arXiv:1107.2847 [hep-th]}}].

\bibitem{Witten}
E.~Witten,
{\it Conformal field theory in four and six dimensions,}
[\href{http://xxx.lanl.gov/abs/0712.0157}
{{\tt arXiv:0712.0157 [math.RT]}}];
{\it Some comments on string dynamics},
In *Los Angeles 1995, Future perspectives in string theory* 501-523
[\href{http://xxx.lanl.gov/abs/hep-th/9507121}
{{\tt arXiv:hep-th/9507121}}].

\bibitem{Wittentwo}
E.~Witten,
{\it On S-duality in abelian gauge theory,}
Selecta Math.\  {\bf 1}, 383 (1995)
[\href{http://xxx.lanl.gov/abs/hep-th/9505186}
{{\tt arXiv:hep-th/9505186}}].

\bibitem{Verlinde}
E.~P.~Verlinde,
{\it Global aspects of electric - magnetic duality,}
Nucl.\ Phys.\ B {\bf 455}, 211 (1995)
[\href{http://xxx.lanl.gov/abs/hep-th/9506011}
{{\tt arXiv:hep-th/9506011}}].

\bibitem{Henningson}
M.~Henningson,
{\it The quantum Hilbert space of a chiral two form in d = (5+1)-dimensions,}
JHEP {\bf 0203}, 021 (2002)
[\href{http://xxx.lanl.gov/abs/hep-th/0111150}
{{\tt arXiv:hep-th/0111150}}].

\bibitem{Dirac}
P~.A.~M.~Dirac,
\underline{Lectures on quantum mechanics},
New York: Belfer Graduate School of Science, Yeshiva University
(1964) 87p.

\bibitem{Das}
A.~Das,
\underline{Lectures on quantum field theory},
Hackensack, USA: World Scientific (2008) 775 p.

\bibitem{AS}
M. Abramowitz and I. Stegun, \underline{Handbook of Mathematical Functions},
New York: Dover Publications (1972), p378.

\bibitem{Seibergone} 
M.~Berkooz, M.~Rozali and N.~Seiberg,
{\it Matrix description of M theory on $T^4$ and $T^5$,}
Phys.\ Lett.\ B {\bf 408}, 105 (1997)
[\href{http://xxx.lanl.gov/abs/hep-th/9704089}
{{\tt arXiv:hep-th/9704089}}].

\bibitem{Seibergtwo} 
N.~Seiberg,
{\it Notes on theories with 16 supercharges,}
Nucl.\ Phys.\ Proc.\ Suppl.\  {\bf 67}, 158 (1998)
[\href{http://xxx.lanl.gov/abs/hep-th/9705117}
{{\tt arXiv:hep-th/9705117}}].

\bibitem{Hull} 
C.~M.~Hull,
{\it BPS supermultiplets in five-dimensions,}
JHEP {\bf 0006}, 019 (2000)
[\href{http://xxx.lanl.gov/abs/hep-th/0004086}
{{\tt arXiv:hep-th/0004086}}].

\bibitem{Bern}
Z.~Bern, J.~J.~Carrasco, L.~J.~Dixon, M.~R.~Douglas, M.~von Hippel and 
H.~Johansson,
{\it D = 5 maximally supersymmetric Yang-Mills theory diverges at six loops,}
Phys.\ Rev.\ D {\bf 87}, 025018 (2013)
[\href{http://lanl.arxiv.org/abs/1210.7709}
{{\tt arXiv:hep-th/1210.7709}}].

\bibitem{Coxeter}
H. Coxeter and W. Moser, \underline{
Generators and Relations for Discrete Groups},
New York: Springer Verlag (1980), p85.

\end{thebibliography}
\end{document}